\documentclass[12pt]{article}
\usepackage{epsfig, epsf, graphicx}
\usepackage{pstricks, pst-node, psfrag}
\usepackage{amssymb,amsmath,bm}
\usepackage{verbatim,enumerate}
\usepackage{enumitem}
\usepackage{rotating, lscape,natbib}
\usepackage{setspace}
\usepackage{subfigure}
\usepackage{tikz}
\usetikzlibrary{arrows,calc,tikzmark}
\usepackage{hyperref}
\usepackage{float}
\usepackage{caption}
\usepackage{natbib}
\usepackage{comment}
\usepackage{dsfont}
\usepackage{booktabs}
\newtheorem{defi}{Definition}

\newtheorem{prop}{Proposition}

\newtheorem{corollary}{Corollary}
\newtheorem{assumption}{Assumption}

\newtheorem{Example}{Example}

\begin{document}

\thispagestyle{empty} \baselineskip=28pt \vskip 5mm
\begin{center} {\Huge{\bf Functional Time Series Analysis Based on Records }}
	
\end{center}

\baselineskip=12pt \vskip 10mm

\begin{center}\large
Israel Mart\'inez-Hern\'andez\footnote[1]{\label{note1} 
\baselineskip=10pt Department of Mathematics $\&$ Statistics, Lancaster University Lancaster, United Kingdom. \\
E-mail: i.martinezhernandez@lancaster.ac.uk }$^{, \ref{note2}}$
 and Marc G.~Genton\footnote[2]{\label{note2} 
\baselineskip=10pt Statistics Program,
King Abdullah University of Science and Technology,
Thuwal, Saudi Arabia.\\
E-mail: marc.genton@kaust.edu.sa \\
This research was supported by the
King Abdullah University of Science and Technology (KAUST). \\
The authors thank the King Abdullah City for Atomic and Renewable Energy (K.A. CARE) for providing the wind speed observational data. }

\end{center}

\baselineskip=17pt \vskip 10mm \centerline{\today} \vskip 15mm

\begin{center}
{\large{\bf Abstract}}
\end{center}

In many phenomena, data are collected on a large scale and of different frequencies. In this context, functional data analysis (FDA) has become an important statistical methodology for analyzing and modeling such data. The approach of FDA is to assume that data are continuous functions and that each continuous function is considered as a single observation. Thus, FDA deals with large-scale and complex data. However, visualization and exploratory data analysis, which is very important in practice, can be challenging due to the complexity of the continuous functions. Here we propose some nonparametric tools for functional data observed over time (functional time series). For that, we propose to use the concept of record. We study the properties of the trajectory of the number of record curves under different scenarios. Also, we propose a unit root test based on the number of records. The trajectory of the number of records over time and the unit root test can be used as visualization and exploratory data analysis. We illustrate the advantages of our proposal through a Monte Carlo simulation study. We also illustrate our method on two different datasets: Annual mortality rates in France and daily wind speed curves at Yanbu, Saudi Arabia. Overall, we can identify the type of functional time series being studied based on the number of record curves observed.

\baselineskip=14pt

\par\vfill\noindent
{\bf Some key words:} Functional time series visualization; Functional depth; Functional unit root test; Non-stationary functional time series; Record curves.
\par\medskip\noindent
{\bf Short title}: Statistical Tools Based on Record

\clearpage\pagebreak\newpage \pagenumbering{arabic}
\baselineskip=26pt

\section{Introduction}\label{sec:intro}


Due to modern technologies, data can now be collected on a large scale and in an automatic fashion for many phenomena, resulting in high-dimensional and high-frequency data, that can be considered as continuous functions or surfaces (images). For example, in economy,  finance, climatology, medicine, biology, and engineering, data can be collected with characteristics that vary along a continuum (time or  space). Functional Data Analysis deals with this type of data, where each continuous function can represent daily or monthly profiles, and these profiles are considered as a single point observation \citep[see, e.g.,][]{Ramsay-Silverman2005}. Here we assume that our data are a functional time series observation, i.e., a sequence of curves observed over time. 

An important part of data analysis is visualization and exploratory data analysis. This helps to decide whether to transform the data or what class of models to use.  In general, visualization and exploratory data analysis provide data characteristics that are not apparent from statistical models. In the context of functional data some of these tools are functional bagplots and functional highest density region plots \citep[][]{HyndmanAndShang2010}, the functional boxplot \citep[][]{SunGenton2011}, and magnitude-shape plot \citep[][]{Dai-Genton2018}. Although these tools are useful, they do not provide characteristics that vary over time. Here, we propose some additional tools using the concept of records. The main advantage of the record concept is that it is  invariant under monotonic transformations of the data. This allows  us to cover a large class of functional time series data. In addition, records can be used to identify extreme or abnormal  curves. With that motivation, we extend the record definition to functional data, and then we obtain nonparametric tools based on this extension.

The record theory has been studied extensively for a sequence $\{W_{1}, \ldots, W_{n}\}$ of identically distributed univariate random variables for both independent and dependent data \citep{Sparre54, Feller71, BalleriniAndResnick87, LindgrenEtAl87, Burridge-Guerre96,  Ahsanullah2015}.  It studies the events that exceed all previous observations, i.e., $W_{n}> \max\{W_{1}, \ldots, W_{n-1}\}$.
 The two most studied quantities of records are the probability for a record at time $n$ and the number of records observed up to time $n$. It is well known that the expected number of records for stationary time series grows at rate $\log n$ \citep[][]{LindgrenEtAl87}. On the other hand, if the time series is a random walk process, the growth rate is $n^{1/2}$  \citep{Sparre54,Feller71,Burridge-Guerre96}. Moreover, if the time series has a linear trend component, then the number of records grows at rate $n$ \cite[][]{BalleriniAndResnick87}. Extensions of the study of records to multivariate data  can be found in the literature \citep[see][]{GoldieAndResnick89, GoldieAndResnick95,Gnedin98, Wergen2012, DombryEtal2018, FalketAt2018}. 

A challenge in the extension of record definition to functional data space is that there is no natural way to define an order on this space. This  makes the definition of minimum and maximum curves problematic.   
 Here, we propose an extension of the definition of records by using an order for functional data based on depth notions. Then, we study the  behavior of the number of functional records (the two most extreme curves) under stationarity and under stochastic trend components. By visualizing the growth rates of the number of functional records, we can infer if the functional time series is or not  stationary, and we can infer on the different type of trends. For a more formal test, we propose a unit root test based on the number of record curves.
 
  Several notions of depth (called functional depth) have been proposed for functional data, including integrated depth \citep[][]{FraimanAndMuniz2001}, 
  band depth and modified band depth  \citep[][]{Lopez-PintadoRomo2009}, half-region depth based on hypographs and  epigraphs \citep{Lopez-PintadoRomo2011}, spatial depth \citep{Chakraborty&Chaudhuri2014} and extremal depth \citep{NarisettyEtAl2016}. Other functional depth definitions can be found in   \cite{Nieto-ReyesEtAl2016}, \cite{GijbelsAndNagy2017}, and \cite{Huang2019}. Depth has been used in different statistical problems. For example to detect outliers, to obtain robust estimators and to define functional boxplots, taking advantage of its center-outwards order  \citep[]{RousseeuwHubert99,FraimanAndMuniz2001, SunGenton2011, SgueraEtal2014, MARTINEZHERNANDEZ2019}. The order induced by the functional depth can be viewed as order statistics. Unlike the usual order statistics in $\mathbb{R}$, ordered from the smallest value to the largest, the order based on depth starts with the most central curve that corresponds to the highest depth value, and moves further away from the center, ending with the most outlying curve that corresponds to the smallest depth value.  We use this center-outwards ordering to define functional records.
  
 
 In this paper, we are interested in studying two applications of functional time series: wind speed curves in Saudi Arabia and mortality rates in France. Let $X_{i}(s)$ be the daily curves of wind speed at $80$m $[m/s]$ where  $i=1, \ldots, n$ represents the day, and $s\in [0, 24)$ represents hours within a day. The study of the wind speed curves is important for renewable energy generations. By using record curves,  we can describe the dynamics of the record daily wind speed. It is relevant to know when and how often a record curve is observed to predict the efficiency of wind turbines and to prevent disruption and possible damage to a wind farm. Moreover, with the information of record curves, we can classify the underlying functional process and then obtain a better predictor.  Now, let  $X_{i}(s)$ denote the mortality rate in year $i$, at age $s$. It is important to know (besides prediction) how these rates behave over the years, taking into account all ages. By studying the functional records, we analyze whether the new functional records over the years correspond to the natural randomness of the process, or if there is an indication of a decreasing trend. In general, the number of functional records provides information about the stationarity and nonstationarity properties of the functional time series. 


The main contributions presented in this paper are: $1)$ the establishment of  a generalized definition of upper and lower record for functional time series;  $2)$ the study of the  growth rate of the number of functional records over time, under stationarity and nonstationarity assumptions; and 
$3)$ the introduction of a unit root test for a general integrated of order one $(I(1))$ functional process, as an application of the functional record. The contributions $2)$ and $3)$ provides tools for visualization and exploratory data analysis. 
    
The remainder of our paper is organized as follows: In Section \ref{Pre}, we introduce mathematical concepts for functional data, functional time series, and functional depth. In Section \ref{URT}, we describe an extension of records to functional data. In Section \ref{PropRecord}, we study the properties of the number of functional records, both for stationary and nonstationary functional time series.  In Section \ref{TP}, we propose a unit root test as an application of the study of functional records. In that section, we conduct a simulation study to evaluate the performance of the proposed test.  In Section \ref{DA}, we illustrate our proposal  on two different datasets: the daily curves of wind speed at Yanbu, Saudi Arabia, and the annual mortality rates for males in France. Section \ref{Dis} presents some discussion. Proofs are provided in the Appendix. 

\section{Preliminaries}\label{Pre}
\subsection{Functional time series}
Throughout this paper, we assume that our data is a collection of $n$ functional observations $\{x_{1}(s), \ldots, x_{n}(s)\}$ with $s\in \mathcal{T}$. Without loss of generality, we assume $\mathcal{T}=[0,1]$.  Let  $X_{i}$ be a functional random variable  defined on a separable Hilbert space $\mathcal{H}$ equipped with an inner product $\langle \cdot , \cdot \rangle$ and a norm $\|\cdot \|_{\mathcal{H}}$. We assume that $\{x_{i}\}$ is a realization of the functional random variables $\{X_{i}\}$.  We denote by $L_{\mathcal{H}}^{2}= \{X \,;\, \mathbb{E} (\| X \|_{\mathcal{H}}^{2} ) < \infty \}$  the set of random variables $X$ of $\mathcal{H}$ with finite second  moment. If $X$ is a functional random variable with distribution $P$, we write $X\sim P$, and it is said to be symmetrically distributed ($P$ is centrally symmetric) about $z\in \mathcal{H}$ if and only if $X-z=- (X-z)$ in distribution.



 Assuming that $\{X_{i}\}\subseteq  L_{\mathcal{H}}^{2}$, the covariance operator at lag $h$ is defined as $C_{X_{i-h},X_{i}}(z)=\mathbb{E}\{ \langle X_{i-h} - \mu_{i-h}, z \rangle (X_{i}- \mu_{i}) \}$,  for all  $z\in \mathcal{H}$, and $\mu_{i}= \mathbb{E}(X_{i})$. Then $X_{i}\in L_{\mathcal{H}}^{2}$ is said to be (weakly)  stationary if (i) $\mathbb{E}(X_{i})=\mu$ for all $i$ and (ii) $C_{X_{i+h}, X_{j+h}}(z)=  C_{X_{i}, X_{j}}(z), \, z\in \mathcal{H}$ for all $h$. If $i=j$ we write $C_{X_{i}}$ instead of $C_{X_{i}, X_{j}}$, and $C_{X_{i}, X_{i+h}}= C_{h}$ for stationary functional time series.  
 We consider the Hilbert-Schmidt norm for covariance operators defined as $\| C_{h} \|_{\mathcal{S}}= \{\int_{\mathcal{T}}\int_{\mathcal{T}} \gamma_{h}^{2}(u,v) \mathrm{d}u \mathrm{d}v \}^{1/2} $, where $\gamma_{h}(u,v)= \mathrm{Cov}\{X_{0}(u), X_{h}(v)\}$.

\begin{assumption}\label{A0}
All functional random variables are defined on a common probability space. We assume the observed functional time series is a realization of a sequence $\{X_{i}\}\subseteq  L_{\mathcal{H}}^{2}$. Also, we assume the mean function and the covariance operator are well defined.
\end{assumption}

We denote by $\mathcal{B}_{\mathcal{H}}$ the space of linear operators from $\mathcal{H}$ to $\mathcal{H}$ and by $\|\cdot \|_{\mathcal{B}_{\mathcal{H}}}$ the corresponding operator norm.
Let $\{\varepsilon_{i}, \, i \in \mathbb{Z}\}$ be an i.i.d. sequence in $L_{\mathcal{H}}^{2}$, and let $\{\Psi_i\} \subseteq \mathcal{B}_{\mathcal{H}}$. A functional linear process $\{X_{i},\, i \in \mathbb{Z}\}$ with innovations $\{\varepsilon_{i}\}$ is defined as 
\begin{equation}\label{LP}
X_{i}(s)= \sum_{j=0}^{\infty} \Psi_j(\varepsilon_{i-j})(s),\, u\in [0,1].
\end{equation}
If  $\sum_{j=0}^{\infty}  \|\Psi_j\|_{\mathcal{B}_{\mathcal{H}}}^2 <\infty$, then the series $\{X_{i}\}$ is convergent in $L_{\mathcal{H}}^{2}$  \citep[][]{Bosq2000}. In this case, the functional linear process is stationary. The long-run covariance operator of the linear process is defined as $V=\Psi C_{\varepsilon_{0}} \Psi^{*}\in \mathcal{B}_{\mathcal{H}}$, where $\Psi = \sum_{j=0}^{\infty}  \Psi_j$, $\Psi^{*}$ is the adjoint of the operator $\Psi$, and $C_{\varepsilon_{0}}$ is the covariance operator of $\varepsilon_{0}$. One of the most popular models for functional time series is the functional autoregressive model of order $p$, FAR$(p)$ \citep[][]{Horvathetal2010, KokoszkaR2013, AueEtAl2015}. FAR$(p)$ processes can be seen as a particular case of a functional linear process \citep[][]{Bosq2000}. We refer to \cite{Ramsay-Silverman2005} and \cite{Bosq2000} for a deeper understanding of functional random variables. 

\subsection{Depth for functional data}

Several notions of functional depth have been proposed. The modified band depth (MBD) is one of the most popular functional depth and has motivated the development of extensions, modifications, and generalizations of functional depth definitions. Let $x\in \mathcal{H}$ and let $\mathbf{x}_{1:n} = \{x_{1}, \ldots, x_{n}\}$ be a sample of $X\sim  P$. The MBD of $x$ with respect to the sample $\mathbf{x}_{1:n} $ computes the proportion of time that the curve $x$ is in a band constructed by two curves from $\mathbf{x}_{1:n} $. Then, the depth value is obtained by averaging the proportion of time over all possible bands. That is,
\begin{equation}\label{MBD}
\mathrm{MBD}(x; \mathbf{x}_{1:n} )= \binom{n}{2}^{-1}\! \sum_{1\leq i_{1} < i_{2}\leq n } \lambda\left[ \{s\in [0,1]\, |\, \min(x_{i_{1}}(s), x_{i_{2}}(s) )\leq x(s)\leq \max(x_{i_{1}}(s), x_{i_{2}}(s) ) \} \right], 
\end{equation}
where $\lambda$ is the Lebesgue measure on $[0,1]$. The corresponding population version is denoted by $\mathrm{MBD}(x; P)$. 
The definition \eqref{MBD} is for a band obtained with two different curves. However,  the band can be obtained by more than two curves \citep[see][for more details]{Lopez-PintadoRomo2009}. 

Another functional depth is the extremal depth (ED). The ED of $x\in \mathcal{H}$ with respect to $\mathbf{x}_{1:n} $ computes the pointwise extremeness of the curve $x$. Namely, let $D_{x}(s;\mathbf{x}_{1:n} ):= 1-  | \sum_{i=1}^{n}[\mathds{1}\{x_{i}(s)<x(s)\}  - \mathds{1}\{x_{i}(s)>x(s)\} ]  | / n $ be the pointwise depth of $x$, taking values in $\mathbb{D}\subset \{0,1/n,\ldots, 1 \}$.  Let $G_{x}(r)= \int_{0}^{1}\mathds{1}\{ D_{x}(s,\mathbf{x}_{1:n} )\leq r\}\mathrm{d}s$, for each $r\in \mathbb{D}$, be the corresponding cumulative distribution function. Let $0\leq d_{1}< d_{2}<\cdots < d_{M} \leq 1$ be the ordered elements of the depth levels obtained from $D_{x}$. Then, $x\curlyeqprec x_{i}$ if $G_{x}(d_{1}) > G_{x_{i}}(d_{1})$, and $x\curlyeqsucc x_{i}$ if $G_{x}(d_{1}) < G_{x_{i}}(d_{1})$. If $G_{x}(d_{1}) = G_{x_{i}}(d_{1})$, then the comparison is based on $d_{2}$ and repeated until the tie is broken. If $G_{x}(d_{j}) = G_{x_{i}}(d_{j})$, for all $j=1,\ldots, M$, then the two functions are equivalent in terms of depth. Finally, the ED of $x$ is defined as 
\begin{equation}\label{ED}
\mathrm{ED}(x; \mathbf{x}_{1:n}) = 1-  \frac{\# \{i : x\curlyeqprec  x_{i}\}}{n },
\end{equation}
where the corresponding population version is denoted by $\mathrm{ED}(x; P)$.
See \cite{NarisettyEtAl2016} for more details. 

In this paper, we do not assume any specific functional depth for the theoretical study, but we require regularity conditions to be satisfied. 
\begin{assumption}\label{AfD}  We assume the functional depth satisfies the conditions of nondegeneracy, maximality at the center, scalar-affine invariance, and monotonicity from the center. 
\end{assumption}
Assumption \ref{AfD} is a common assumption required for functional depth definition.  For a discussion of the above conditions and other functional depth definitions, see \cite{Nieto-ReyesEtAl2016} and \cite{GijbelsAndNagy2017}. Throughout this paper, a functional depth refers to a functional depth satisfying Assumption \ref{AfD} . For illustration purposes, we will use the depths \eqref{MBD} and \eqref{ED} implemented in the \textit{fda} \citep[][]{fda} and \textit{fdaoutlier} \citep[][]{fdaoutlier} R  \citep[][]{R} packages, respectively.

\section{Definition of Functional Records}\label{URT}
One of the challenges for functional data (as well as for multivariate data) is that there is no natural way to define an order on this space. Here, we propose to extend the definition of records by using an order for functional data based on depth notions. As a motivation, we first introduce the record definition for univariate scalar time series.

\subsection{Classical records}
Let $\{W_1, \ldots, W_n\}$ be a sequence of continuous random variables in $\mathbb{R}$ (observe that $W_{i}=W_{j}$ with probability zero for $i \neq j$). Records are defined for times $t=2, \ldots, n$ and involves comparing the values up to this time (the first observation can always be defined as reocord). Let $W_{(1)}, \ldots,W_{(t)}$ be the corresponding order statistics for the first $t$ random variables. The random variable $W_t$ is defined as an upper record if $W_t=W_{(t)}$, and a lower record if $W_t=W_{(1)}$, with probability one \citep{Ahsanullah2015}. Finally, $W_t$ is a record if it is a lower or upper record. When a depth notion is used, a center-outwards ordering is obtained, $W_{[1]}\leq W_{[2]}\leq \cdots \leq W_{[t]}$ with probability one, where $W_{[i]}$ is the random variable with the $i$th largest depth value among the $t$ random variables. In this case, $W_{[t]}$ and $W_{[t-1]}$ are the two most extreme observations. Under certain conditions on the depth definition, the set of the smallest and the largest order statistics  is equal to the set of the two most extreme observations, i.e., $\{W_{(1)}, W_{(t)}\}=\{W_{[t-1]},W_{[t]}\}$. Therefore, the classical records and the extreme observations identified with depth notions can be equivalent.  Based on these observations, we extend the classical records to a functional record definition, and we study the behavior of the number of functional records. 

\subsection{Functional records using depth}
Suppose we observe a functional time series $\{ x_{1},\ldots, x_{n}\}$, with distribution $P$. For $t=2, \ldots, n$, let $\mathbf{x}_{1:t}= \{ x_{1},\ldots, x_{t}\}$ be the first $t$ curves. For $1\leq i, j, \leq t$, let $\mathrm{fD}_{i,t}:=\mathrm{fD}(x_{i}; \mathbf{x}_{1:t})$ be the value of the estimated functional depth. We define an order ``$ \prec  $'' among the curves  as  $x_{i} \prec x_{j}$ if $\mathrm{fD}_{i,t}>\mathrm{fD}_{j,t}$ and we say that $x_{j}$ is more extreme than $x_{i}$. If $\mathrm{fD}_{i,t}=\mathrm{fD}_{j,t}$, i.e., if there are ties, we say that $x_{i}$ and $x_{j}$ are equally deep (extreme), and we use the notation $x_{i} \sim x_{j}$. Let $x_{[i],t}$ denote the curve corresponding to the $i$th largest depth value in $\mathbf{x}_{1:t}$. Then, $x_{[1],t}, \ldots, x_{[t],t}$ can be viewed as order statistics, with $x_{[1],t}$ representing the deepest curve and $x_{[t],t}$ the most outlying curve. The order statistics induced by depth start with the most central curve that corresponds to the biggest $\mathrm{fD}_{i,t}$ value, and move further away from the center, ending with the most extreme curve that corresponds to the smallest $\mathrm{fD}_{i,t}$ value. 
\begin{figure}[!b]
\begin{center} 
 \includegraphics[scale=.42]{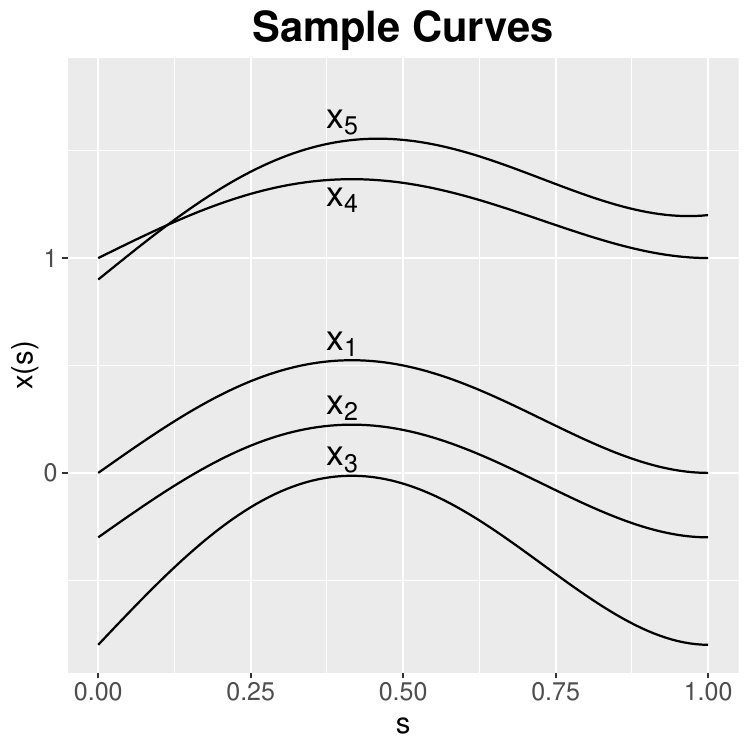}
 \includegraphics[scale=.42]{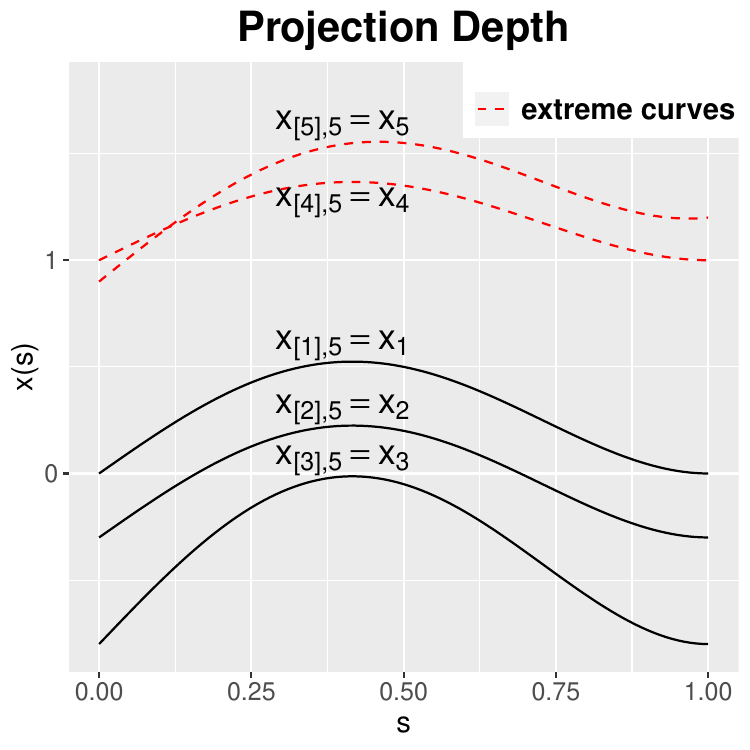}
 \includegraphics[scale=.42]{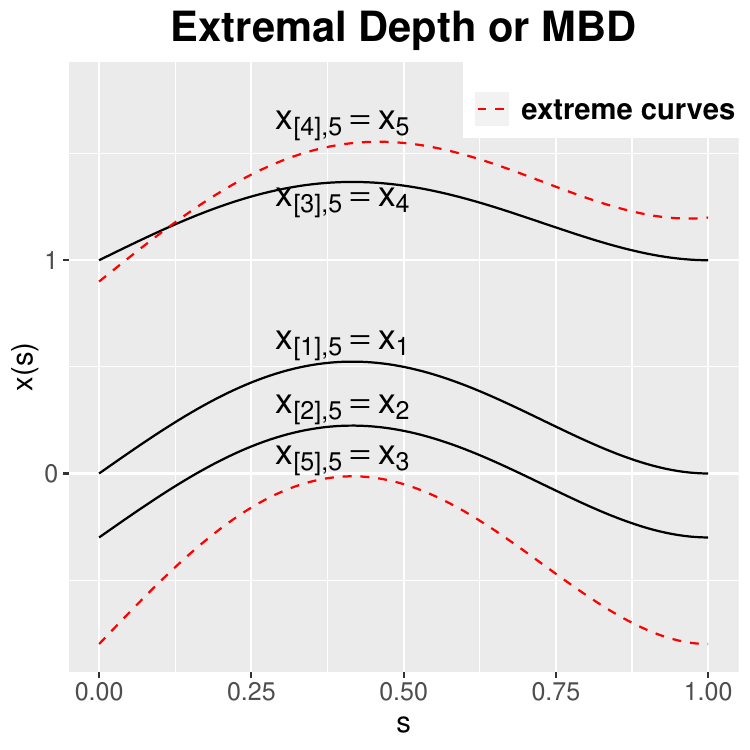}
\caption{Two different results in the ordering of curves. Dashed curves represent the two most extreme curves. When ordering is induced by  functional projection depth, the two most extreme curves are $x_{[4],5}= x_{4}$ and $x_{[5],5}=x_{5}$. While when ordering is induced by extremal depth or MBD, the two most extreme curves are $x_{[4],5} =x_{5}$ and $x_{[5],5}=x_{3}$.}
\label{Assumption1}
\end{center}
\end{figure}

However, not all functional depth definitions imply that $x_{[t-1],t}$ and $x_{[t],t}$ are the most extreme curves in terms of record definition, i.e., ``upper''  and ``lower'' curves. For example, let us consider $t=5$ with curves as in Figure \ref{Assumption1}, and let us assume it is a sample functions of $X\sim P$. Let $\mathrm{fD}(x;P)= \{1+ o(x; P) \}^{-1} $ be the projection functional depth \citep[][]{Zuo2003}, where $o(x; P)= \int_{0}^{1}  \frac{| x(s)-\mathrm{median}(X(s)) | }{\mathrm{MAD}(X(s)) } \mathrm{d}s $ is the integrated Stahel-Donoho outlyingness \citep[][]{Stahel81, Donoho82}. After computing $\mathrm{fD}_{1,5}, \ldots, \mathrm{fD}_{5,5}$, the ordering induced by this functional depth is $x_{1} \prec x_{2}\prec x_{3} \prec x_{4} \prec x_{5} $, with $ x_{4}$ and $x_{5}$ being the two most outlying curves. At time $t=5$, the curve $x_{5}$  clearly can be considered as a record curve but not $x_{4}$. Now, if $\mathrm{fD}(x;P)$ is the extremal depth or MBD, the ordering induced is $x_{1} \prec x_{2}\prec x_{4} \prec x_{5} \prec x_{3} $, with $ x_{3}$ and $x_{5}$ being the two most outlying curves. In this case, at $t=5$,  $x_{3}$ can be definitely classified as an extreme curve as well as $x_{5}$, which makes more sense in the context of record definition. 
To avoid functional depths such as the projection functional depth, we impose a condition on $\mathrm{fD}$ in a set of  constant functions:
\begin{assumption}\label{A1} Let $a_{1}, \ldots, a_{n}$ be a sequence of any real numbers with $a_{i}\neq a_{j}$ for $i\neq j$, and let $\{x_{i}(s):=a_{i}\mathds{1}_{[0,1]}(s), t=1, \ldots,n, \, s\in \mathcal{T}\}$ be a sequence of functions. The functional depth $\mathrm{fD}$ satisfies $\{a_{(1)}\mathds{1}_{[0,1]}(s), a_{(n)}\mathds{1}_{[0,1]}(s)\}= \{x_{[n],n} (s), x_{[n-1],n } (s)\}$ as a set, for all $s\in \mathcal{T}$, where $a_{(i)}$ is the usual order statistics. 
\end{assumption}
Some examples of functional depth definitions that satisfy Assumption \ref{A1} are MBD, extremal depth, and spatial depth. 

In the illustration above with $\mathrm{fD}$ as extremal functional depth (Figure \ref{Assumption1}), the last extreme curve observed among the five curves corresponds to $x_{5}$ (an upper functional record), but it does not correspond to the smallest functional depth value. The smallest functional depth value corresponds to the curve $x_{3}$ (a lower functional record). In general, a new functional record does not always correspond to the smallest functional depth value,  but it is one of the two smallest.  
\begin{defi} \label{RecordDefinition}
Let $\mathbf{x}_{1:t}= \{ x_{1} , \ldots,  x_{t}\}$ be an observed functional time series up to time $t$, $t\geq 2$. Let $\mathrm{fD}$ be a functional depth satisfying Assumptions \ref{AfD} and \ref{A1}. $x_{t}$ is called a functional record at time $t$ if $$\mathrm{fD}_{t,t}\in \{\mathrm{fD}_{(t),t}, \mathrm{fD}_{(t-1),t}\},$$ 
where $\mathrm{fD}_{(i),t}$ denotes the $i$th largest value of the functional depths $\mathrm{fD}_{1,t}, \ldots ,\mathrm{fD}_{t,t}$, and $\mathrm{fD}_{i,t}= \mathrm{fD} (x_{i};\mathbf{x}_{1:t} )$.
\end{defi}

The  nondegeneracy of functional depth is important in the functional record definition, because if the functional depth definitions suffer a degeneracy problem, i.e., with probability one the depth value is zero for every function in a general class of continuous Gaussian processes  \citep{ChakrabortyChaudhuri2014b}, then each function can be a functional record.

We observe that ties do not affect Definition \ref{RecordDefinition} unless it occurs with the smallest and second smallest functional depth values, $\mathrm{fD}_{(t-1),t}$ and $\mathrm{fD}_{(t),t}$ (See Assumption \ref{A2} in the next section). Let $x_{j}$ be the last functional record observed at time $t-1$. According to Definition \ref{RecordDefinition}, if $\mathrm{fD}_{t,t}= \mathrm{fD}_{j,t}$, then $x_{t}$ is a new functional record at time $t$. This makes sense since $x_{t}(s)= x_{j}(s)$ with probability zero if $P_{X(s)}$ is a continuous distribution function. Thus, unlike the classical record definition in $\mathbb{R}$, we define  $x_{t}$ as a functional record if it is  equally extreme as the previous two most extreme curves at time $t-1$. 


	
A functional record $x_{t}$ at time $t$ may be visually easy to classify as upper or lower functional records. One way to  define upper and lower functional record is using the deepest curve (median curve) as a reference curve, and computing the proportion $T^{u}_{t}$ of time that $x_{t}$ is above the median curve and the proportion $T^{l}_{t}$ of time that $x_{t}$ is below the median curve. 
\begin{defi}\label{URecordD} Let $x_{t}$ be a functional record  at time $t$ according to Definition \ref{RecordDefinition}. We say that $x_{t}$ is an upper record if $T^{u}_{t}:= \int_{0}^{1} \!  \mathds{1}\{x_{t}(s) > x_{[1],t}(s)  \} \mathrm{d}s > T^{l}_{t}:=\int_{0}^{1} \!  \mathds{1}\{x_{t}(s) < x_{[1],t}(s)  \} \mathrm{d}s $, and a lower record in the  other case.
\end{defi}

In some scenarios, it is possible that $T^{u}_{t}= T^{l}_{t}$. Here, we assume that such ties do not occur. 

\begin{assumption}\label{notT} Let $x_{t}$ be a functional record  at time $t$ according to Definition \ref{RecordDefinition}.  We assume that $P(T^{u}_{t}= T^{l}_{t})=0$. 
\end{assumption}
 
\begin{figure}[!t]
\begin{center} 
 \includegraphics[scale=.41]{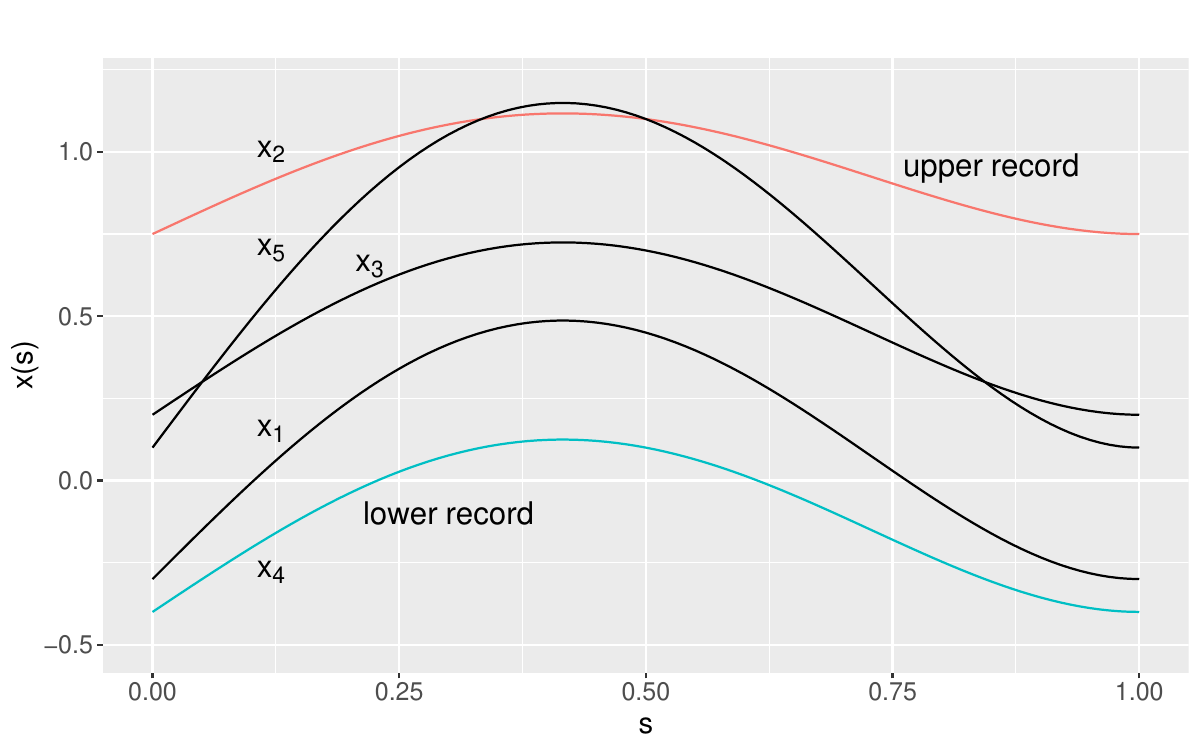}
 \includegraphics[scale=.41]{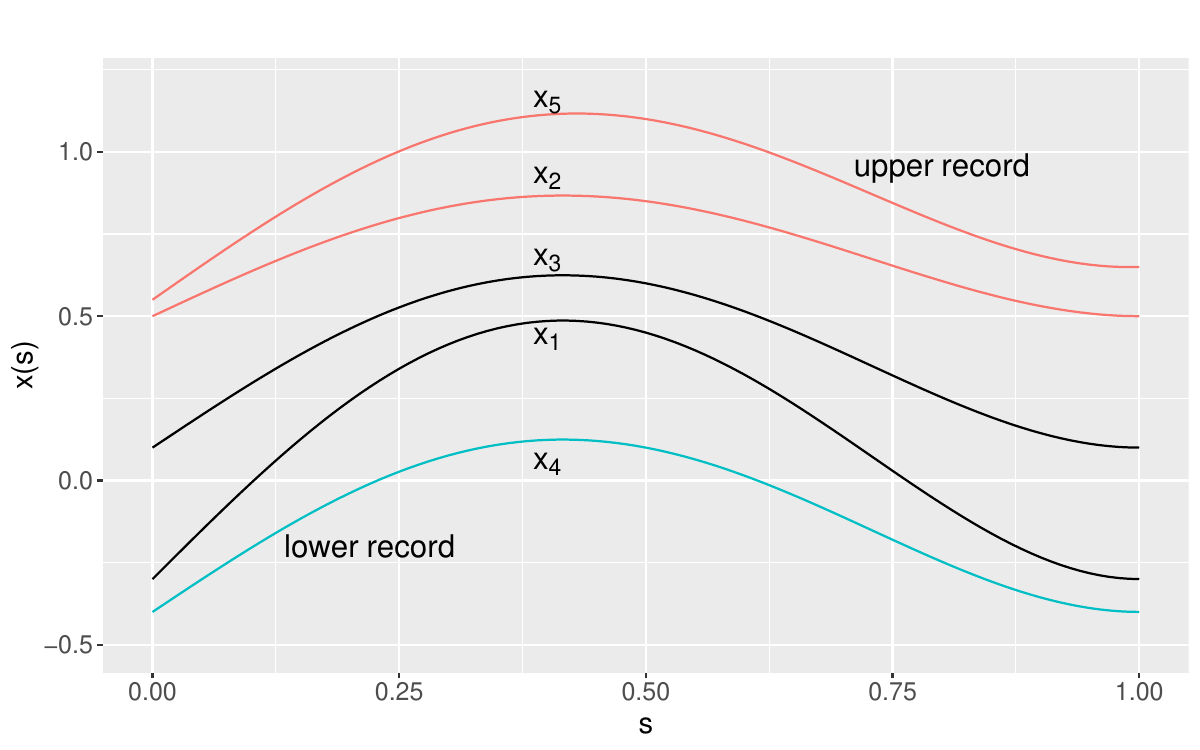}
\caption{Functional records with $n=5$ curves. The upper functional record is indicated by the red curve, and the lower functional record is indicated by the blue curve. Left: functional records are observed  at $t=2$ and $t=4$. Right: functional records are observed at $t=2$, $t=4$, and $t=5$.}
\label{ExDef1} 
\end{center}
\end{figure}
Figure \ref{ExDef1} shows an example of functional record observed over time with sample size $n=5$. Here, we use MBD to
compute the functional records (a similar result is obtained with extremal depth). In the left plot, we obtain two functional records, at $t=2$ and $t=4$, corresponding to the upper functional record and the lower functional record, respectively. The curve $x_{4}$ is the last functional record observed at time $t=5$ since $x_{5}$ is deeper than $x_{2}$ and $x_{4}$ in the sample.  In contrast, in the right plot, we obtain three functional records observed at $t=2$, $t=4$, and $t=5$, where $x_{2}$ and $x_{5}$ are upper functional records, and $x_{4}$ is a lower functional record. In this case, $x_{5}$ is the last functional record observed, since $x_{2}$ is now deeper than $x_{4}$ and $x_{5}$ in the sample. In this last case, we obtain ties in the last two extreme curves, i.e., $ \mathrm{fD}_{4,5} =  \mathrm{fD}_{5,5}$. However, these curves are on opposite sides of the sample, hence it makes sense to define $x_{5}$ as a new functional record. 


\section{Properties of Functional Record Number}\label{PropRecord}


 Let $\mathbf{X}=\{X_{i}(s), i\geq 1  \}$ be a sequence of functional random variables with distribution $P$. Let $\mathrm{fD}$ be a functional depth that satisfies Assumptions \ref{AfD} and \ref{A1}. For $t\geq 2$, let $X_{[1],t}, \ldots ,X_{[t],t}$ be the order statistics induced by the functional depth $\mathrm{fD}(X_{1}; \mathbf{X}_{1:t}), \ldots, \mathrm{fD}(X_{t}; \mathbf{X}_{1:t})$, where $ \mathbf{X}_{1:t}$ is the first $t$ functional random variables.  Then, $X_{t}$ is defined as a functional record at time $t$ if 
 \begin{equation}
 X_{t} \in \{X_{[t-1],t},X_{[t],t}\}.
 \end{equation}
 For $i=1, \ldots, t$, we define $T^{u}_{i,t}$ as the proportion of time at which $X_{i}$ is above the central curve $X_{[1],t}$,  $T^{u}_{i,t}:=\int_{0}^{1} \!  \mathds{1}\{ X_{i}(s) > X_{[1],t}(s)  \} \mathrm{d}s$,  and $T^{l}_{i,t}:=\int_{0}^{1} \!  \mathds{1}\{ X_{i}(s) < X_{[1],t}(s)  \} \mathrm{d}s$ as the proportion of time at which  $X_{i}$ is below $X_{[1],t}$.  If $X_{t}$ is a functional record at time $t$, we say that $X_{t}$ is an upper functional record if $T_{t,t}^{u}>T_{t,t}^{l}$, and a lower functional record otherwise.

 We study the number of functional records over time.  Let $R_{t}=\mathds{1} \{X_{t}\,\, \mathrm{ is \, a\, functional\, record}\}$ be the indicator of $X_{t}$ being a functional record at time $t$, and let $N_{t}$ be the counting process representing the number of functional records up to time $t$, i.e.,
 \begin{equation}
N_{t}=\sum_{i=1}^{t} R_{i},
 \end{equation}
 where $R_{1}:=1$. We define the functional record times as $L(1)=1$, $L(2)=2$, and for $ k=3,4,\dots $,
$L(k)=\min\{t \,:\, t>L(k-1) \mbox{ and }  R_{t}=1  \}.$  We use the notations $R_{t}^{u}, N_{t}^{u}$, and $L^{u}(k)$ to denote the respective variables for the upper functional records. Notice that the definition of  $L^{u}(k)$ is such that  the events $\{N_{t}^{u} \geq k \}$ and $\{ L^{u}(k)\leq t \}$ are equivalent (similarly with $L(k)$).

With  Assumption \ref{notT}, we have that a lower functional record is an upper functional record of the process $\{-X_{i}\}$. Therefore, we focus on the upper functional records. To establish the theoretical properties of the process $N^{u}_{t}$, we assume the following:
\begin{assumption}\label{A2} Let $\mathbf{X}_{1:t}=\{X_{1}, X_{2}, \ldots, X_{t}\}$ be a sequence of functional random variables, with $t\geq 3$. If $R_{t}=1$, then, with probability one 
\begin{align*}
\max & \{\mathrm{fD}(X_{t};\mathbf{X}_{1:t} ), \mathrm{fD}(X_{L^{u}(N^{u}_{t-1})}; \mathbf{X}_{1:t}), \mathrm{fD}(X_{L^{l}(N^{l}_{t-1})}; \mathbf{X}_{1:t})\}< \\
&\quad  \min \{\mathrm{fD}(X_{i}; \mathbf{X}_{1:t});\, i \in \{1,\ldots,t-1\}\backslash \{L^{u}(N^{u}_{t-1}),L^{l}(N^{l}_{t-1})\} \}.
\end{align*}
\end{assumption}
Assumption \ref{A2} means that ties of depth values are allowed, but if $X_{t}$ is a functional record at time $t$ then $\mathrm{fD}(X_{t};\mathbf{X}_{1:t} )$ can only tie with $\mathrm{fD}(X_{L^{u}(N^{u}_{t-1} )};\mathbf{X}_{1:t} )$ or  $\mathrm{fD}(X_{L^{l}(N^{l}_{t-1})};\mathbf{X}_{1:t} )$. Also, with Assumption \ref{A2}, it is not possible to observe more than one upper (lower) functional record at time $t$. In general, it is unlikely to observe ties in the depth values, i.e., that $\mathrm{fD}_{i,t}= \mathrm{fD}_{j,t}$ for some $i,j\in \{1, \ldots, t\} $  and $i\neq j$. However, it does not have a zero probability, especially for small sample sizes and functional depth taking values  of the form $1/j$. Although, ties can be broken by using an auxiliary sequence of i.i.d. random variables $W_{i}$, $i=1, \ldots, t$, such that $W_{1}$ has an absolutely continuous distribution and independent of $P$ \citep[see, e.g.,][]{Dufour2006}. Then a strict and total order can be obtained as follows: $(x_{i}, W_{i}) \prec (x_{j}, W_{j})$ if and only if $\mathrm{fD}_{i,t} > \mathrm{fD}_{j,t}$ or if $\mathrm{fD}_{i,t} = \mathrm{fD}_{j,t}$ and $W_{i}>W_{j}$. Thus, Assumption \ref{A2} is not restrictive.

In the univariate case, it is known that if the time series is an independent sequence or a stationary time series satisfying the Berman condition, then $N^{u}_{t}$ grows at rate $\log t$ \citep{LindgrenEtAl87}. On the other hand, if the time series is a random walk  process, then the growth rate of $N^{u}_{t}$ is $t^{1/2}$ \citep{Sparre54,Feller71,Burridge-Guerre96}. With the previous definitions, we show similar results for functional records. 

We observe that, if $\{X_i, i\geq 1\}$ is an independent sequence of functional random variables, then $P(R^{u}_{t}=1)= 1/t$ for any ranking definition. Indeed, the probability of $X_{t}$ being a record is the probability of $X_{t}$ taking a specific place among $\{1,\ldots, t\}$. Then, $N_{t}^{u}= O(\log t)$ with probability one.
\begin{prop}\label{EstacionaryCase}
 Let $\{X_i\}$ be a stationary functional time series  such that $\log(h) \| C_{h}\|_{{\mathcal{S}} }\to 0$ as $h\to \infty$. 
 Let ``$\prec$'' be an ordering such that $P( X_{i} \prec X_{j}\,\, \mathrm{or} \,\, X_{j} \prec X_{i})=1$ for all pairs $i\neq j$.  Then, under Assumptions \ref{A0}--\ref{notT} $$\lim_{t \to \infty} \frac{N^{u}_{t}}{\log t}= O(1),$$
with probability one.
\end{prop}
Proof: See Appendix. 


The condition on the covariance operator in Proposition \ref{EstacionaryCase}  is not restrictive for functional time series, and it holds  if the functional time series is $L^{2}$-$m$-approximable \citep[][]{HormanKokoszka}.  \cite{HormanKokoszka} showed that this approximation is  valid for linear and non-linear functional time series. In particular, the FAR$(1)$ model with coefficient operator that has norm less than one is $L^{2}$-$m$-approximable.


\begin{figure}[!t]
\begin{center} 
\subfigure[]{\includegraphics[scale=.33]{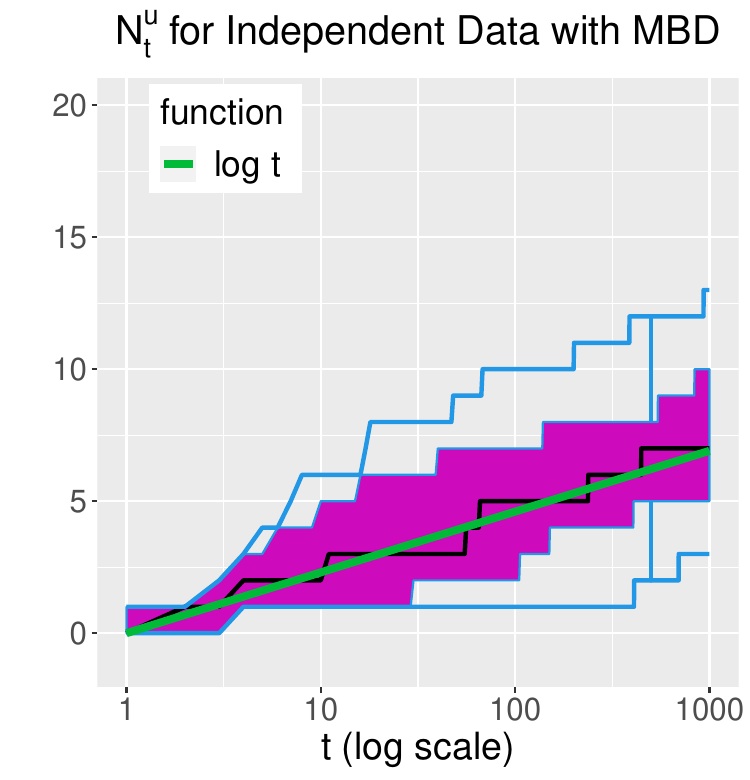}}
\subfigure[]{\includegraphics[scale=.33]{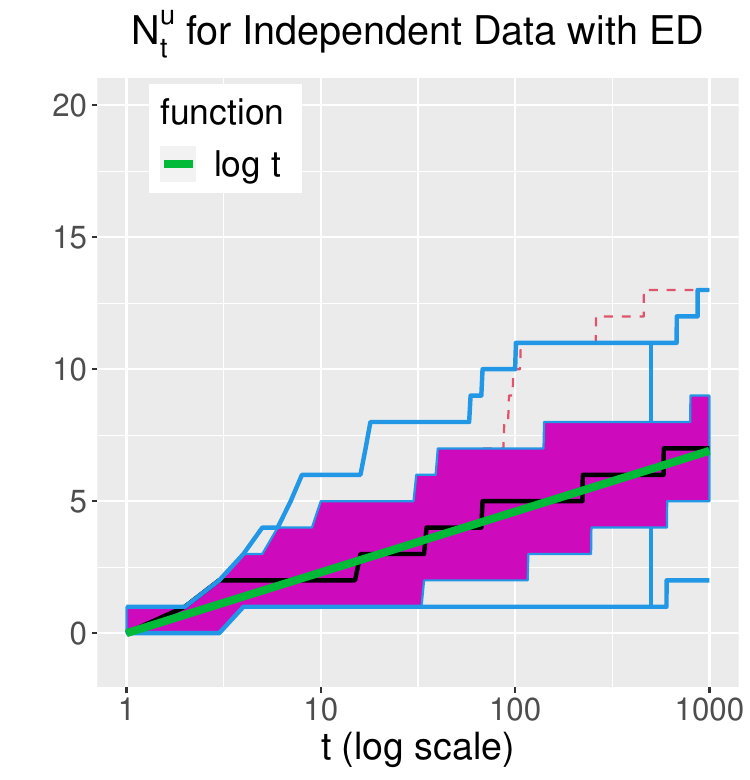}}
\subfigure[]{\includegraphics[scale=.33]{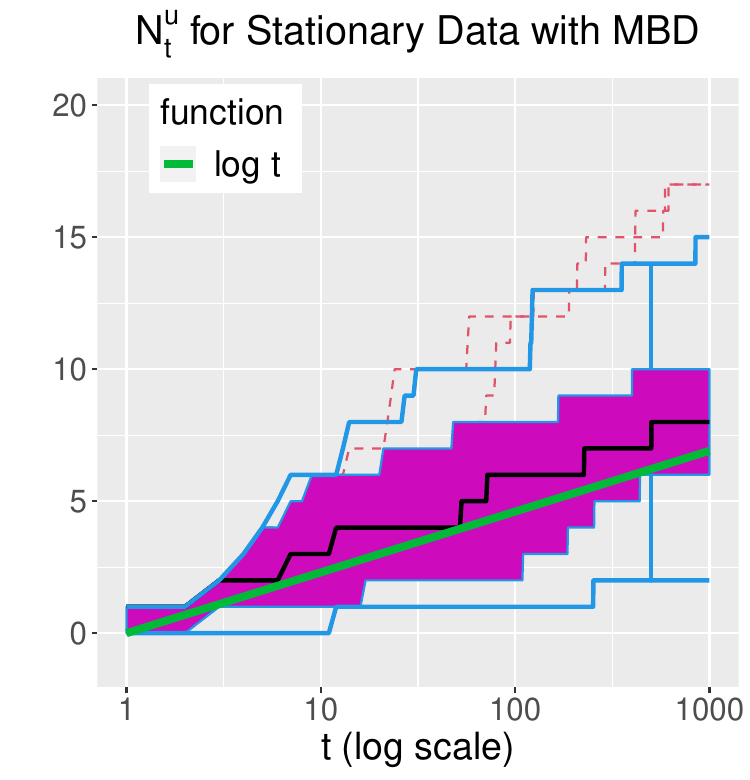}}
\subfigure[]{\includegraphics[scale=.33]{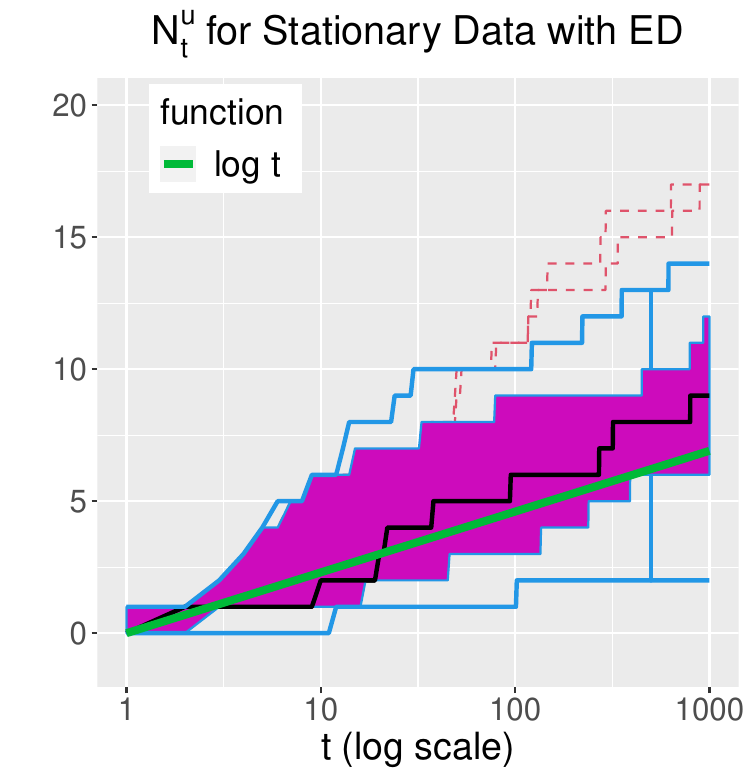}}
\caption{Log scale functional boxplot of $100$ trajectories of $N_{t}^{u}$ by using MBD and ED with $t=2,\ldots, 1000$. Each trajectory of $N_{t}^{u}$ is obtained from $\{X_{i}\}_{i=1}^{n}$ where $n=1000$, and $\{X_{i}\}$ is an independent functional sequence  (a,b) and a stationary functional sequence (c,d). The green curve represents the $\log t$ function.}
\label{BoxplotStationary} 
\end{center}
\end{figure}

We simulate $X_{i}=\varepsilon_{i}, i=1,\ldots, n=1000$ as an independent sequence, where, for each $i$, $\varepsilon_{i}$ is a Brownian motion in $[0,1]$. Figure \ref{BoxplotStationary}(a) and \ref{BoxplotStationary}(b) show the functional boxplot \citep[][]{SunGenton2011} of $100$ trajectories of $N_{t}^{u}$ with $t=2, \ldots, n$, using MBD and ED on the independent sample curves. In Figure \ref{BoxplotStationary}(c) and  \ref{BoxplotStationary}(d), we simulate stationary functional time series from $X_{i}(s)= c_{1}\int_{0}^{1}\!\beta(u,s) X_{i-1}(u) \mathrm{d}u + \varepsilon_{i}(s)$, where 
$\beta(u,s)= \exp\{- (u^{2} + s^{2} )/2 \},$ and $c_{1}$ is such that $ \left \{\int_{0}^{1} \int_{0}^{1} c_{1}^{2} \beta(u,s)^{2}\mathrm{d}u\mathrm{d}s\right \}^{1/2}=0.5$. We observe that  $N_{t}^{u}$ has the same growth rate in all cases, i.e., $\log t$.

Now, we state the result for values of   $N_{n}^{u}$ under a nonstationary functional process.
\begin{prop} \label{LimDistiUR}
 Let $X_i= X_{i-1} + \varepsilon_{i}$ be a functional random walk with $\{\varepsilon_{i}\}$ an i.i.d. sequence in $L^{2}_{\mathcal{H}}$. Under Assumptions \ref{A0}--\ref{A2}, and if $\varepsilon_{0}$ has a symmetric distribution about the mean, we have that
 \begin{equation}\label{UpperAD}
  \frac{N^{u}_n}{\sqrt{n}}  \overset{ d }{\longrightarrow}  G_{1},
 \end{equation}
when $n \rightarrow \infty$, where $G_{1}$ is a random variable with probability density function  $g_{1}(u)= \frac{1}{\sqrt{ \pi}}  \exp{(- u^{2}/4 )}$ for $u\geq 0$.
\end{prop}
Proof: See Appendix. 

\begin{figure}[!t]
\begin{center} 
  \includegraphics[scale=.33]{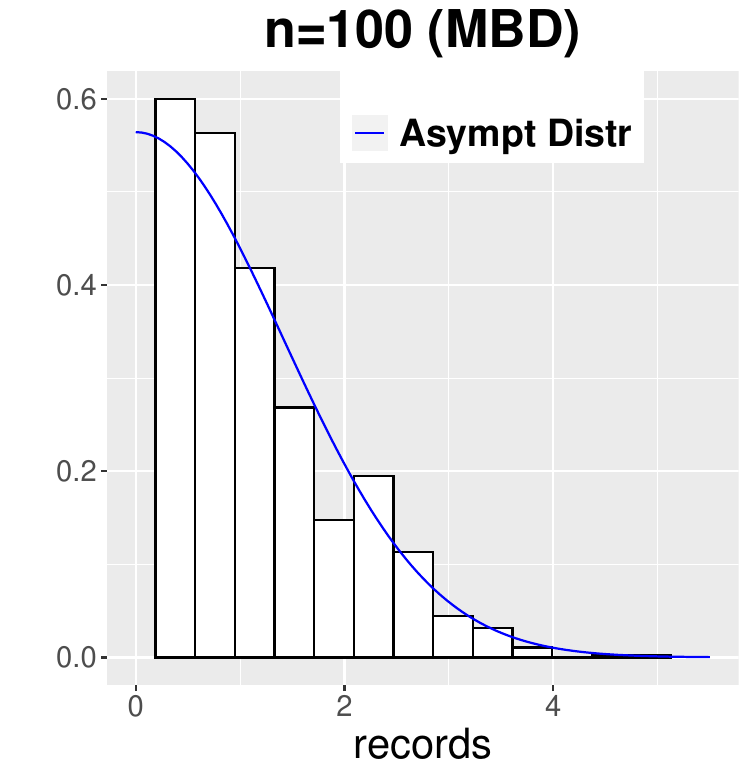}
  \includegraphics[scale=.33]{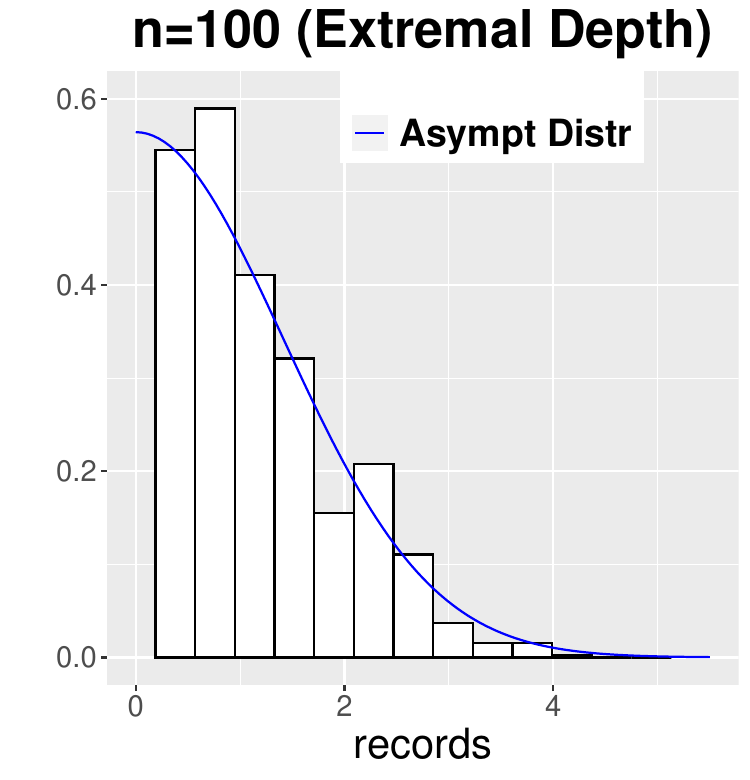}
  \includegraphics[scale=.33]{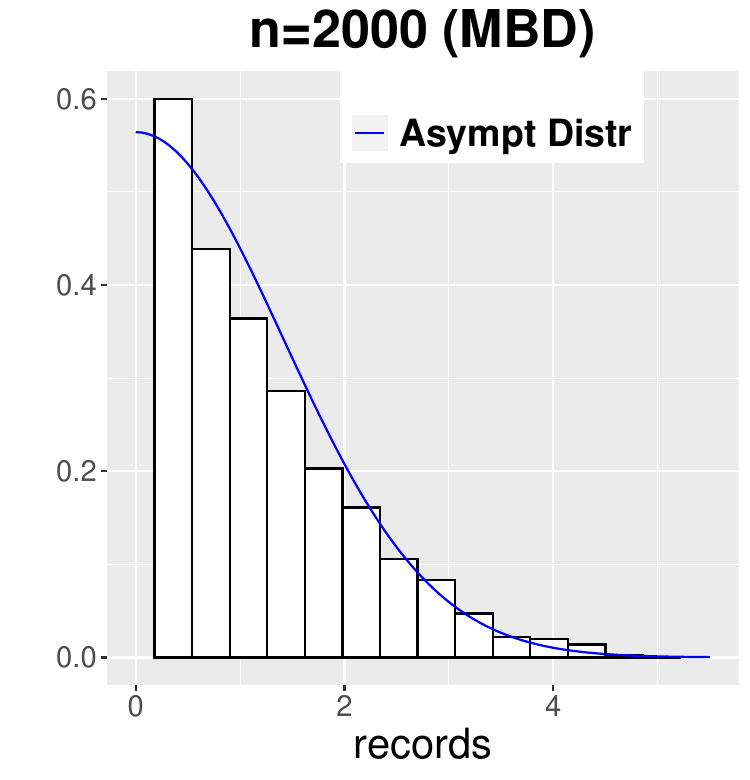}
  \includegraphics[scale=.33]{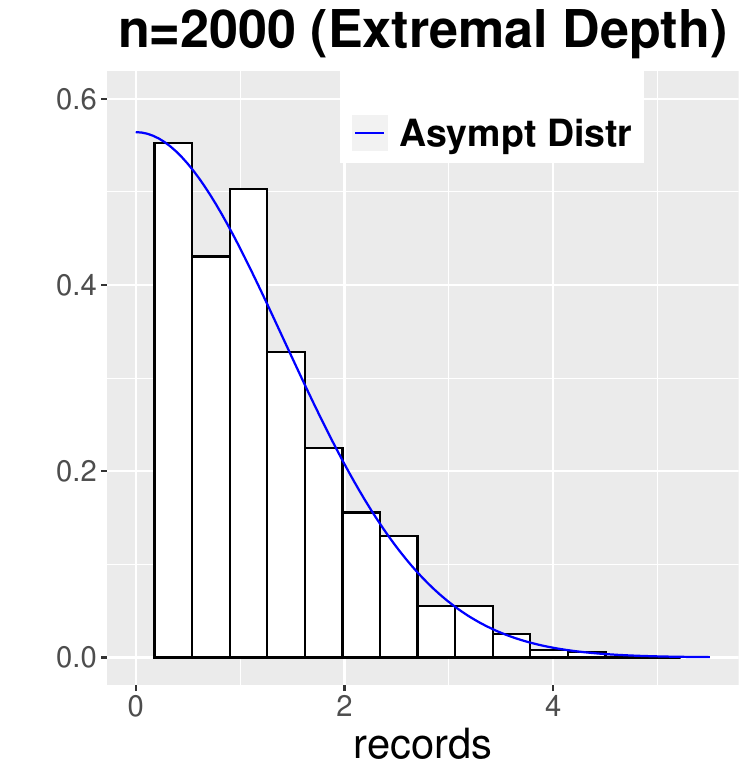}
\caption{Histogram of values of $N^{u}_{n}/\sqrt{n}$ with $n=100,2000$, and the asymptotic distribution (solid blue curve) from Proposition \ref{LimDistiUR}.}
\label{Emp-Asympt} 
\end{center}
\end{figure}

Proposition \ref{LimDistiUR} can be generalized to $I(1)$ functional processes (see Section \ref{TP}).  As an illustration of the result in Proposition \ref{LimDistiUR}, we simulate a functional random walk, with Brownian motion in $[0,1]$, as a functional white noise, and for different sample sizes  $n=100$, and $n=2000$ . We simulate $1000$ replicates of each case, and then obtain $1000$ replicates of values of $N_{n}^u $.  Figure \ref{Emp-Asympt} shows histograms of $N_{n}^{u}/ \sqrt{n}$ where the solid blue curve represents the asymptotic distribution from Proposition \ref{LimDistiUR}.  We observe that the asymptotic distribution provides a better description of the empirical distribution when the sample size increases, particularly with MBD. However, with a sample size $n=100$, this approximation is already reasonably good.

  
\section{Application to Functional Unit Root Test} \label{TP}
Records have been used in different problems, in particular to test for a unit root \citep[see][]{Burridge-Guerre96,AparicioEtAl2006}.  
In this section, we propose a unit root test for functional time series that uses the normalized counting process $N_{n}=N_{n}^{u}+N_{n}^{l}$. One advantage of using records to test for a unit root is that it is a nonparametric test; it is also robust  against structural breaks, and it does not involve the estimation of any coefficient operators that could be a difficult task and therefore face computational issues. Moreover, a  unit root test based on records is invariant under monotonic transformations of the data, since records are invariant too. Thus, we can use records to test a unit root in a general class of $I(1)$ functional processes.

\subsection{$I(1)$ functional processes}
We assume a general definition of the $I(1)$ functional process that involves the Johansen approach. 
 Assume that $\{X_i\}$ is a functional linear process such that the first difference $\Delta X_i:=X_{i}- X_{i-1}$ admits the functional linear representation in \eqref{LP} with innovations $\{\varepsilon_{i}\}$, i.e.,
$\Delta X_i= \sum_{j=0}^{\infty} \Phi_j(\varepsilon_{i-j}),$ 
where $\{\Phi_j \}\in \mathcal{B}_{\mathcal{H}}$, and $\sum_{j=0}^{\infty} j \|\Phi_j\|_{\mathcal{B}_{\mathcal{H}}} <\infty$. Let $C_{\varepsilon_0}$ be the covariance operator of $\varepsilon_{0}$ that is positive definite, and denote by  $\Lambda=\Phi C_{\varepsilon_0} \Phi^{*}$ the long-run covariance operator of $\{\Delta X_{i}, \, i\geq 1\}$, with $\Phi= \sum_{i=0}^{\infty} \Phi_i$. Then, $\{X_{i}, \,i\geq 0\}$ can be written as 
\begin{equation}\label{SolI1}
X_{i}(s)=Z_{0}(s) + \Phi \left( \sum_{j=1}^{i}\varepsilon_{j} \right)(s) + \eta_{i}(s),
\end{equation}
where $Z_{0}\in L^{2}_{\mathcal{H}} $,  and $\{\eta_{i}\}$ is a stationary process in $\mathcal{H}$. The solution \eqref{SolI1} contains an initial condition $Z_{0}$, a functional random walk component $ \Phi(\sum_{j=1}^{i}\varepsilon_{j} )$, and a stationary component $\eta_{i}$. Observe that the functional  random walk component is defined only on $\mathrm{ran}\, \Phi=\{\Phi(z)\,:\, z\in \mathcal{H} \} $, thus the functional time series is stationary on the complementary space. Then, $\{X_i\}$ is an $I(1)$ functional process if, and only if, $\Lambda\neq 0$  \citep[]{BeareEtal2017}. In this paper, we adopt this last definition.
\begin{Example}[FAR$(1)$] \label{Ex1}
Assume that $\{X_{i}\}$ is a FAR$(1)$ process, $X_{i}(s)= \rho_{1}(X_{i-1})(s)+ \varepsilon_{i}(s)$.
\begin{enumerate}
\item If $\rho_{1}= \mathrm{Id}_{\mathcal{B}_{\mathcal{H}}}$, where $\mathrm{Id}_{\mathcal{B}_{\mathcal{H}}}$ denotes the identity operator from $\mathcal{H}$ to $\mathcal{H}$, then $X_{i}$ is a functional random walk and can be written as 
$$X_{i}(s)= X_{0}(s) +\sum_{j=0}^{i-1} \varepsilon_{i-j}(s).$$
Therefore $\{X_{i}\}$ is an $I(1)$ functional process.
\item Suppose $\rho_{1}\neq\mathrm{Id}_{\mathcal{B}_{\mathcal{H}}}$ has  one eigenvalue equal to one, that is, there exist $v\in \mathcal{H}$ such that $\rho_{1}(v)=v$, and the operator pencil $A(z)=\mathrm{Id}_{\mathcal{H}} -z \rho_{1}$ is not invertible at $z=1$. Moreover, assume the space $\mathcal{H}$ can be decomposed as $\mathcal{H}=\mathrm{ran}\, A(1) \oplus \mathrm{ker}\,A(1)$. Then, $X_{i}$ can be written as in \eqref{SolI1}, and therefore $\{X_{i}\}$ is an $I(1)$ functional process. In this case, $\{\langle X_{i}, z \rangle\}$ is a univariate $I(1)$ process for all $z\notin \mathrm{ker} \,A(1)$,  and $\{\langle X_{i}, z \rangle\}$ is stationary, up to a choice of the initial condition, for all $z\in \mathrm{ker}\, A(1)$ \citep[see][]{BeareEtal2017}. 
\end{enumerate}
\end{Example}
Additional discussions about functional processes with a unit root can be found in  \cite{MassimoandPaolo2017} and \cite{BeareEtal2020}. 

Under the FAR$(1)$ model (or FAR$(p)$ model), the presence of a unit root affects the accuracy of the estimation of the coefficient operator $\rho_{1}$ because then most  of the properties of existing estimators will not hold, and we need to consider an alternative model.  Therefore, a functional unit root has an impact on both estimation and modeling. Thus, we need to detect the unit root accurately. 

\subsection{Functional records for $I(1)$ processes}

Loosely speaking, an $I(1)$ functional process has two components: a functional random walk component and a stationarity functional process component. The trajectory of the functional random walk component leads the trajectory of the $I(1)$ functional process. Thus, the number of records for the $I(1)$ functional process will be similar to that of a functional random walk. We formalize this result in the following proposition.

\begin{prop} \label{LimDist3}
 Let $\{X_i\}$ be an $I(1)$ functional process in $\mathcal{H}$.  Assume $\{\varepsilon_{i}\}$ is an i.i.d. sequence in $L_{\mathcal{H}}^{2}$ with symmetric distribution about the mean, and Assumptions \ref{A0}--\ref{A2} are satisfied. Then, the  corresponding normalized random variable $N_{n}^u/ \sqrt{n}$ has the same asymptotic distribution as the corresponding one for a functional random walk in Proposition \ref{LimDistiUR}.
\end{prop}
Proof: See Appendix.  

Next we defined a record-based (RB) functional unit root test.  
We  consider the testing of the null hypothesis of an $I(1)$ functional process versus the functional process is stationary. In other terms,
$$H_{0}: \{X_{i}\}\,\, \mathrm{ is\,\, an } \,\,I(1)\,\,  \mathrm{ functional\,\, process }\,  \mbox{ vs } \, H_{1}: \{X_{i}\} \,\, \mathrm{ is\,\, a \,\, stationary \,\, process },$$
where the corresponding innovations $\{\varepsilon_{i}\}$ are assumed to be symmetrically distributed. For sample size $n$, the test statistic $T_{n}$ for the RB-functional unit root test is the number of upper and lower records normalized with $\sqrt{n}$, i.e.,  $T_{n}= n^{-1/2} N_{n} =n^{-1/2}  ( N^u_{n} + N^l_{n} )$. 
 \begin{corollary}\label{AD}
 Let $\{X_i\}_{i=1}^{n}$ be a realization of a functional linear process with innovations $\{\varepsilon_{i}\}$ having a symmetric distribution about the mean. Under Assumptions \ref{A0}--\ref{A2}, we have that
 \begin{enumerate}
 \item under the null hypothesis, $T_{n} \overset{d}{\longrightarrow} G_{2},$
 where $G_{2}$ is a random variable with probability density function $g_{2}(u)=\sqrt{\frac{2}{\pi}} \, u^{2} \exp{(- u^{2}/2 )}$, $u\geq 0$, and 
 \item under the alternative hypothesis, $T_{n} \overset{p}{\longrightarrow} 0$.
 \end{enumerate}
 \end{corollary}
 Proof: See Appendix.
 
From Corollary \ref{AD}, we use the left tail of the asymptotic distribution of the test statistic $T_{n}$  to  test for a functional unit root, i.e., given the significance level  $\alpha$, reject $H_{0}$ if $T_{n}$ is smaller than the quantile $q_{\alpha}$ of $g_{2}(u)$, where $q_{\alpha}$ is the quantile of order $\alpha$ of  $g_{2}(u)$.
 
 In the following sections, we present a Monte Carlo simulation study to evaluate the performance of the test for a finite sample size.

\subsection{Simulation design} \label{MC}

We study the performance of the unit root test, based on functional records under the null  and alternative hypothesis.  We simulate different functional time series, $\{X_{i}(s)\}_{i=1}^{n}$, at $50$ points equispaced on $[0,1]$ with the different sample sizes $n=200, 300, 500$ and $1000$. 
Let $\{\varepsilon_{i}(s)\}$ be a sequence of independent functional random variables. We consider the following models:
 
\begin{enumerate}
\item \label{H01} $X_{i}(s)=X_{i-1}(s) + \varepsilon_{i}(s)$;
\item \label{H02} $X_{i}(s)=\rho(X_{i-1})(s) + \epsilon_{i}(s)$,   where $\rho(z)(s)=a(\langle z , e_{1}\rangle + \langle z , e_{2}\rangle ) e_{1}(s) + a \langle z , e_{1}\rangle e_{2}(s),$ $\{e_{1}, e_{2}\}$ is an orthonormal basis function, and $a= -1/2 + \sqrt{5}/2$.  Assuming that the white noise $\{\epsilon_{i}\}$ satisfies $\mathbb{E}(\langle\epsilon_{i}, e_{1} \rangle^{2} )>0$ but $\mathbb{E}(\langle\epsilon_{i}, e_{2} \rangle^{2} )=0$;
\item \label{H11} $X_{i}(s)= \varepsilon_{i}(s) $;
\item \label{H12} $X_{i}(s)=  \Psi_{1}(X_{i-1})(s) + \varepsilon_{i}(s)$, where  $\Psi_{1}(z)= c_{1} \int_{0}^1  \! \exp\{ (u^2 + s ^2)/2\} z(u)\mathrm{d}u$ and $c_{1}$ is such that $\|\Psi_{1} \|_{ \mathcal{B}_{\mathcal{H}} }=0.5$;
 \item \label{H1C1} $X_{i}(s)= \mu_{1}(s) \mathds{1}_{\{i \leq k\}}  +\mu_{2}(s) \mathds{1}_{\{i > k\}}+ \eta_{i}(s) $, where $\eta_{i}(s)$ is a stationary FAR$(1)$ process as in Model \ref{H12}, $\mu_{1}(s)=0$, $\mu_{2}(s)=2$ and $k=n/2$; and
 \item \label{H1C2} $X_{i}(s)= (\Psi_{1} \mathds{1}_{\{i> k\}}  +\Psi_{2} \mathds{1}_{\{i \leq k\}}) (X_{i-1})(s)+ \varepsilon_{i}(s)$, where $\Psi_{1}$ is as in Model \ref{H12} and $\Psi_{2}(z)= c_{2} \int_{0}^1  \! \exp\{ -(u^2 + s ^2)/2\} z(u)\mathrm{d}u$ where $c_{2}$ is such that $\| \Psi_{2} \|_{ \mathcal{B}_{\mathcal{H}} }= 0.7$, and $k=n/2$.
\end{enumerate}
The choice of the parameter $a$ in Model \ref{H02} makes $\rho$ an operator with an eigenvalue equal to one and the rest has modulus less than one. Models \ref{H01} and \ref{H02} are under the null hypothesis, i.e., $\{X_{i}\}$ is an $I(1)$ functional process, whereas, in Models \ref{H11} to \ref{H1C2}, $\{X_{i}\}$ is not an $I(1)$ functional process. Particularly, in Models \ref{H11} and \ref{H12}, $\{X_{i}\}$ is stationary. We consider the functional white noise to be the Brownian motion (Bm), $\varepsilon_{i}(s)=W_{i}(s), s\in [0,1]$, the Brownian bridge (Bb), $\varepsilon_{i}(s)=W_{i}(s)- s W_{i}(1), s\in [0,1]$, and  $\varepsilon_{i}(s)$ as a stochastic Gaussian process (Gp$(0, \gamma)$) with zero mean and covariance function $\gamma(s,u)= 0.2 \exp\{-0.3 | s-u | \}$ in $[0,1]$.

For comparison, we adopt the regression approach to mimic the Dickey-Fuller test \citep[][]{Dickey-Fuller79}. We fit the model $X_{i}(s)=\int_{0}^{1}\! \beta(u,s)X_{i-1}(u)\mathrm{d}u + \omega_{i}(s) $ for each simulated functional time series, where $ \beta(t,s)$ is estimated using a penalized least square estimator \citep[][]{MARTINEZHERNANDEZ2019}. We then compute the corresponding norm of the coefficient operator, $\{\int_{0}^{1}\int_{0}^{1} \! \beta^{2}(u,s)\mathrm{d}u\mathrm{d}s \}^{1/2}$. If $\{X_{i}\}$ is an $I(1)$ functional process, we expect the norm to be close to one, and if $\{X_{i}\}$ is stationary, we expect the norm to be smaller than one. We report the mean norm values over replicates.

\subsection{Empirical size and power of the test} 
\begin{table}[t!] \centering
\caption{ Empirical size. Proportion of rejections under the null hypothesis. Functional records are obtained using two different functional depths: MBD and ED. We simulate functional time series from Models \ref{H01} and \ref{H02} with different functional white noises, Brownian motion (Bm), Brownian bridge (Bb), and Gaussian process with zero mean and  covariance function $\gamma$ (Gp$(0,\gamma)$). The sample sizes considered are $n=200, 300, 500$ and $1000$. Each scenario is replicated $1000$ times. The values in parentheses indicate the mean value of the norm of the coefficient operator in the FAR$(1)$ model, and they are the same  for both fD. Nominal level is $5\%$. }
\begin{tabular}{lcccccccccc}
\toprule 
 &  &\mbox{Model }\ref{H01}  &   &  &  &     &  \mbox{Model }\ref{H02}  & &  &     \\   \cmidrule{1-1}   \cmidrule{3-6}  \cmidrule{8-11} 

$n$ & &$200$  & $300$  &$500$ &$1000$ &    & $200$& $300$ & $500$ & $1000$      \\  \toprule  \toprule 
$ \mbox{fD=MBD} $ &    &  &   &    &        &      &       &  &  &  \\ 
\toprule 
$\varepsilon_{i}$ &    &  &   &    &        &      &       &  &  &  \\ 
Bm &    &  0.016 & 0.022 &0.017 &     0.029      &      & 0.011&0.015&0.015&0.015 \\
     &    &  (1.61)&(1.66)&(1.65)& (2.13)  & & (1.24)&(1.66)&(1.72)&(1.90)\\ 
Bb &    & 0.002  & 0.003 & 0.009 & 0.015      &      &   -    & -  & - & - \\
    &    & (1.71)&(1.81)&(1.68)&(1.83)& & (-) & (-) & (-) & (-) \\ 
Gp$(0,\gamma)$ &    & 0.034 &0.044 &0.43&0.051     &      &   -  & - & - & - \\
     &    & (1.77)&(1.53)&(1.83)&(1.75) & & (-) & (-) & (-) & (-) \\  \toprule  \toprule 
     $ \mbox{fD=ED} $ &    &  &   &    &        &      &       &  &  &  \\   \toprule 
  $\varepsilon_{i}$ &    &  &   &    &        &      &       &  &  &  \\    
  Bm &    &  0.013 & 0.011 & 0.020 &    0.025   &      & 0.025 & 0.031 & 0.024 & 0.028  \\
  Bb &    & 0.002  & 0.002 & 0.008 & 0.011    &      &   -    & -  & - & - \\
  Gp$(0,\gamma)$ &    & 0.015 &0.023 &0.028 &0.041     &      &   -  & - & - & - \\
\toprule 
\end{tabular}
\label{TableSize}
\end{table}  

We compute the test statistic $T_{n}$ by using MBD and ED as functional depth, and we compare it with the quantile $q_{0.05}$ obtained from the asymptotic distribution in Corollary \ref{AD}.  Table \ref{TableSize} presents the proportion of rejections when the functional time series is under the null hypothesis. 
 We observe that, for Model \ref{H01} with white noise Bm and depth MBD, the proportion of rejection is $0.016$ when $n=200$, and it increases to $0.029$ when $n=1000$.  We observe similar results  when the white noise is Bb. This suggests a slow rate of convergence to the left tail of the asymptotic distribution (for both functional depths). In general, the proportion of rejections gets closer to the chosen significance level as the sample size increases.  In contrast, when the functional white noise is Gp$(0,\gamma)$, we observe a faster convergence of the proportion of rejections to the significance level for both functional depths.  For Model \ref{H02}, there is only one white noise because of the restriction on the model. In this case, when using MBD,  the proportion of rejections is similar to that in Model \ref{H01} with white noise Bb. Whereas when using ED, the proportion of rejections is similar to that in Model \ref{H01} with white noise Bm. With respect to the norm values indicated in parentheses, we observe mean values bigger than one, in all cases. This means, the fitted FAR$(1)$ model is a nonstationary functional time series, in agreement with the data generating processes.
 
\begin{table}[t!] \centering
\caption{ Empirical power. Proportion of rejections under the alternative hypothesis.  Functional records are obtained using two different functional depths: MBD and ED. Functional time series are simulated from Models  \ref{H11} and  \ref{H12}  with different functional white noises, Brownian motion (Bm), Brownian bridge (Bb), and Gaussian process with zero mean and  covariance function $\gamma$ (Gp$(0,\gamma)$). The sample sizes considered are $n=200, 300, 500$ and $1000$. Each scenario is replicated $100$ times. The values in parentheses indicate the mean value of the norm of the coefficient operator in the FAR$(1)$ model, and they are the same  for both fD.}
\begin{tabular}{lcccccccccc}
\toprule 
 &  &\mbox{Model }\ref{H11}   &   &  &  &     &  \mbox{Model }\ref{H12}  & &  &     \\   \cmidrule{1-1}   \cmidrule{3-6}  \cmidrule{8-11} 

$n$ & &$200$  & $300$  &$500$ &$1000$ &    & $200$& $300$ & $500$ & $1000$      \\  \toprule  \toprule 
\mbox{fD=MBD} &    &  &   &    &        &      &       &  &  &  \\   \toprule 
$\varepsilon_{i}$ &    &  &   &    &        &      &       &  &  &  \\ 
Bm &    &0.98&0.99&1.00&1.00     &      &  0.87&0.99&1.00&1.00  \\
  &    &(0.13)&(0.12)&(0.10)&(0.09) & & (0.48)&(0.49)&(0.50)&(0.50) \\ 
Bb &    &  0.92&0.98&1.00&1.00   &      &  0.88&0.99&1.00&1.00\\
  &    & (0.12)&(0.10)&(0.09)&(0.07) & & (0.46)&(0.47)&(0.48)&(0.49) \\ 
Gp$(0,\gamma)$ &    &  0.96&1.00&1.00&1.00     &      & 0.91&0.97&1.00&1.00    \\
  &    & (0.15)&(0.14)&(0.13)&(0.10)& &(0.50)&(0.52)&(0.52)&(0.52) \\   \toprule  \toprule 
  \mbox{fD=ED} &    &  &   &    &        &      &       &  &  &  \\   \toprule 
  $\varepsilon_{i}$ &    &  &   &    &        &      &       &  &  &  \\ 
  Bm &    &0.96&1.00&1.00&1.00     &      &  0.92&0.97&1.00&1.00  \\
Bb &    &  0.90&0.98&1.00&1.00   &      &  0.90&0.98&1.00&1.00\\
Gp$(0,\gamma)$ &    &  0.97&0.99&1.00&1.00     &      & 0.93&0.99&1.00&1.00    \\  
  \toprule 
\end{tabular}
\label{TablePower}
\end{table}  

Our next step is to study the power of the test. Table \ref{TablePower} presents the proportion of rejections under the alternative. Note that Model \ref{H11} represents a stationary, independent sequence of functional data, whereas Model \ref{H12} represents stationary,  dependent functional data. In Model~\ref{H11}, the proportion of rejections is bigger than $0.90$ for small sample sizes, independently of the selection of the white noise $\varepsilon_{i}$ and the functional depth. In Model \ref{H12}, the proportion of rejections is bigger than $0.97$ for sample sizes bigger than $n=300$. In general, the test shows a high power, even for the smaller sample size, $n=200$.    
For norm values, we observe that the respective means of the norms for Models \ref{H11} and \ref{H12} are approximately 
$0.1$ and $0.5$, for all cases. This means that the fitted FAR$(1)$ model is a stationary functional time series, and that the mean norm agrees with the norm of the data generating processes. 

\begin{figure}[t!]
\begin{center}
  \includegraphics[scale=.46]{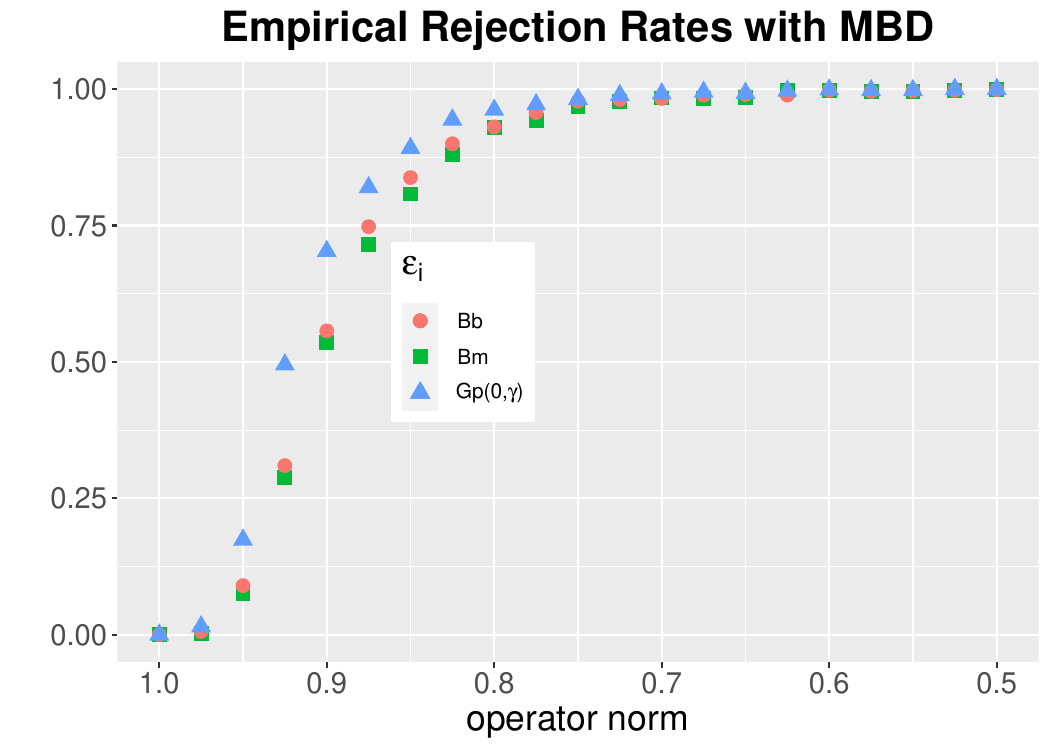}
  \includegraphics[scale=.46]{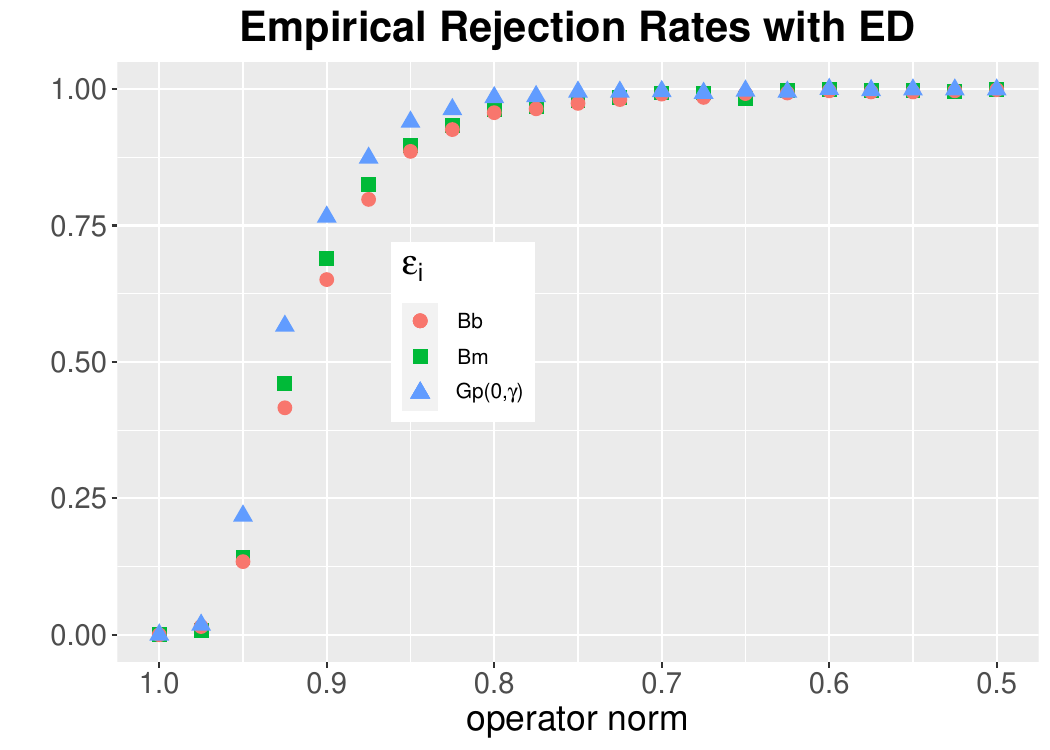}
\caption{Rejection rate for different operator norms of the coefficient operator in Model \ref{H12}. Left: Functional records are obtained with MBD. Right: Functional records are obtained with ED. }
\label{EPowerMod4} 
\end{center}
\end{figure}

Also, we investigate the power curve for Model \ref{H12} when we vary the operator norm $\|\Psi_{1} \|_{ \mathcal{B}_{\mathcal{H}} }=0.5, 0.525,\ldots,0.975 ,1$. Figure \ref{EPowerMod4} shows the rejection rate at level $\alpha=0.05$ with $n=500$ for each different operator norm. We observe that the test has good power, correctly rejecting the null hypothesis when operator norms are smaller than $0.9$. 

As a conclusion, the RB-functional unit root test shows an accurate balance of the significance level and power of the test, already for a medium sample size ($n=300$).  

\subsection{Robustness against structural changes}

One of the advantages of using functional records in the hypothesis test is the robustness to different nonstationary models.  Models  \ref{H1C1} and \ref{H1C2} represent unstable time series, with change on the mean and change on the coefficient operator, respectively. However, Models \ref{H1C1} and \ref{H1C2} are not $I(1)$ functional processes, so we expect to reject $H_{0}$. 
The counting processes $N_{t}$ for Models \ref{H1C1} and \ref{H1C2} should grow at the same rate as that in the stationary case: $N_{t}= O(\log t)$.  Table \ref{TablePower2} shows the corresponding proportion of rejections for these models.

\begin{table}[!t] \centering
\caption{Proportion of rejections against models with structural changes. Functional records are obtained using two different functional depths: MBD and ED. Functional time series are simulated from Models \ref{H1C1} and  \ref{H1C2}  with different functional white noises, Brownian motion (Bm), Brownian bridge (Bb), and Gaussian process with zero mean and covariance function $\gamma$ (Gp$(0,\gamma)$). The sample sizes considered are $n=200, 300, 500$ and $1000$. Each scenario is replicated $1000$ times. The values in parentheses indicate the mean value of the norm of the coefficient operator in the FAR$(1)$ model, and they are the same  for both fD. }
\begin{tabular}{lcccccccccc}
\toprule 
 &  &\mbox{Model }\ref{H1C1}  &   &  &  &     &  \mbox{Model }\ref{H1C2}  & &  &     \\   \cmidrule{1-1}   \cmidrule{3-6}  \cmidrule{8-11} 

$n$ & &$200$  & $300$  &$500$ &$1000$ &    & $200$& $300$ & $500$ & $1000$      \\  \toprule \toprule 
\mbox{fD=MBD} &    &  &   &    &        &      &       &  &  &  \\  \toprule 
$\varepsilon_{i}$ &    &  &   &    &        &      &       &  &  &  \\ 
Bm &    & 0.18&0.43&0.58&0.97     &      &  0.70&0.88&0.94&1.00 \\
 &    & (0.99)&(1.00)&(1.01)&(1.02) & &(0.58)&(0.60)&(0.61)&(0.63) \\ 
Bb &    & 0.19&0.34&0.61&0.98    &      &  0.75&0.91&0.99&1.00 \\
 &    & (0.95)&(0.96)&(0.96)&(0.96) & &(0.57)&(0.59)&(0.60)&(0.61)\\ 
Gp$(0,\gamma)$  &    &0.47&0.73&0.92&1.00       &      &    0.73&0.94&0.99&1.00  \\ 
 &    & (0.95)&(0.96)&(0.97)&(0.98) & &(0.62)&(0.63)&(0.64)&(0.64) \\  \toprule \toprule 
 \mbox{fD=ED} &    &  &   &    &        &      &       &  &  &  \\  \toprule
 $\varepsilon_{i}$ &    &  &   &    &        &      &       &  &  &  \\ 
 Bm &    & 0.34&0.59&0.88&1.00     &      &  0.82&0.96&0.97&1.00 \\
 Bb &    & 0.33&0.54&0.77&0.99    &      &  0.73&0.93&1.00&1.00 \\
 Gp$(0,\gamma)$  &    &0.53&0.86&0.95&1.00       &      &    0.83&0.96&1.00&1.00  \\ 
  \toprule 
\end{tabular}
\label{TablePower2}
\end{table} 
 
For Model \ref{H1C1}, we observe a low proportion of rejections when the sample size is smaller. In this scenario, for a reasonable power, the test requires a sample size bigger than $500$ when using MBD and a sample size bigger than $300$ when using ED. 
 For Model \ref{H1C2}, the results are different. For $n=200$, we observe that the proportion of rejections of the null hypothesis are  bigger than $0.7$ for all white noises and using MBD. Whereas when using ED, this proportion is bigger than $0.82$. We obtain a proportion of rejections bigger than $0.88$, when $n\geq 300$ in all cases. In general, the RB-functional unit root test is robust against structural changes, although a bigger sample size is needed when changes occur on the mean. In contrast, we observe that the norm of the coefficient operator is affected by structural changes. In particular, for Model  \ref{H1C1}, we obtain a mean value of the norms close to one, indicating the possible existence of a unit root on the  data generating processes. In general, we conclude that a test based on regression will have low power in the presence of a structural change, similarly to what occurs with univariate time series.

Our test does not depend on a specific model, and is invariant under monotonic transformations. It is expected to have a good performance with a broad class of models, and in practice, the computation of the number of functional records does not depend on the depth definition.

\section{Data Applications} \label{DA}
In this section, we apply the different tools described in this paper in two different datasets. First, we consider daily curves of the hourly wind speed taken at Yanbu, Saudi Arabia. Our second example involves the annual mortality rates in France (from the R package \textit{demography}, \cite{demography}), from $1816$ to $2006$. We consider the MBD and the ED to compute the functional record curves.

\subsection{Wind speed in Saudi Arabia}
The dataset consists of $n=755$ daily curves of wind speed, at Yanbu, Saudi Arabia, from August 30, $2014$ to September 22, $2016$. Each point of the curve represents wind speed at $80$m $[m/s]$. The study of the behavior of the wind speed is important for renewable  energy generations. Particularly,  by knowing when and how often a record curve of wind speed is observed, we can describe the dynamics of the extreme wind speed curves. An accurate characterization of the extreme daily curves is crucial to predict the efficiencies of wind turbines and energy storage in the presence of an extreme event.



\begin{figure}[!t]
\begin{center}
 \includegraphics[scale=.3]{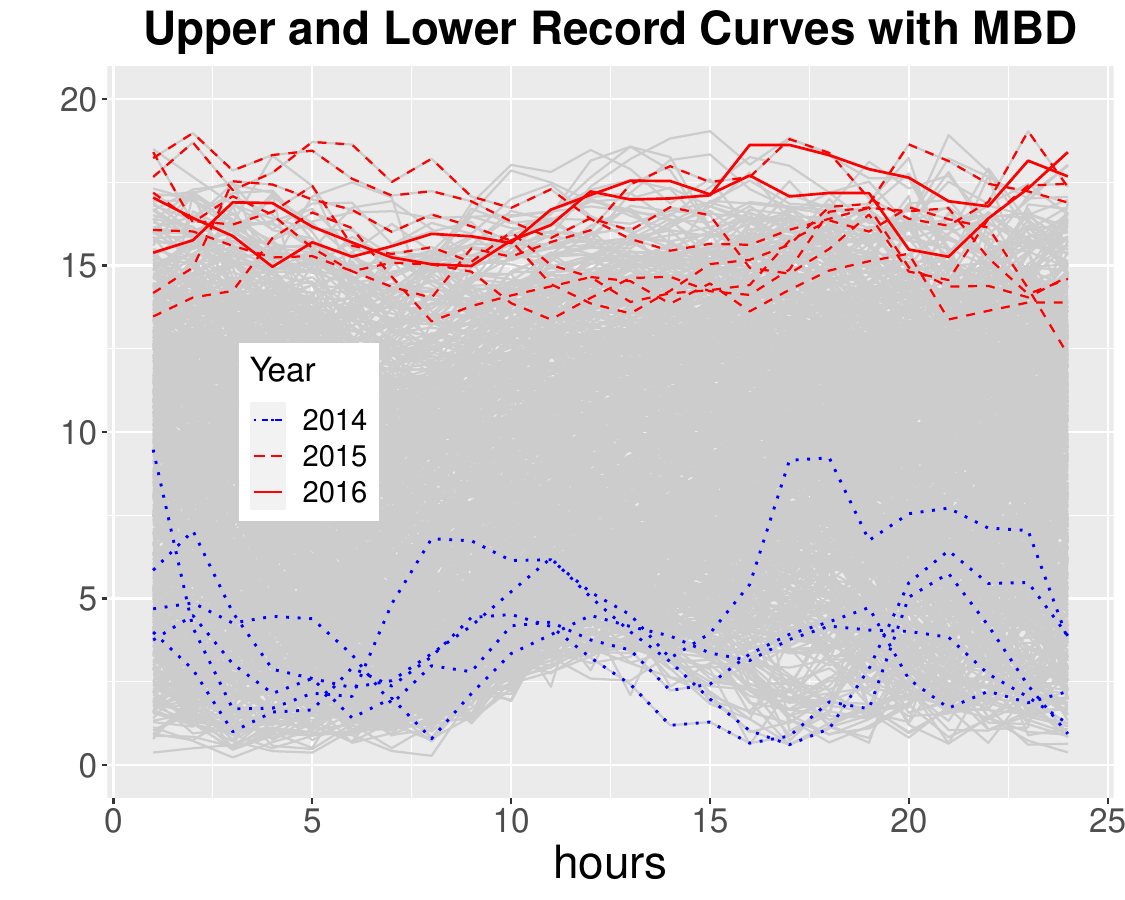}\hspace{-.3cm}
  \includegraphics[scale=.3]{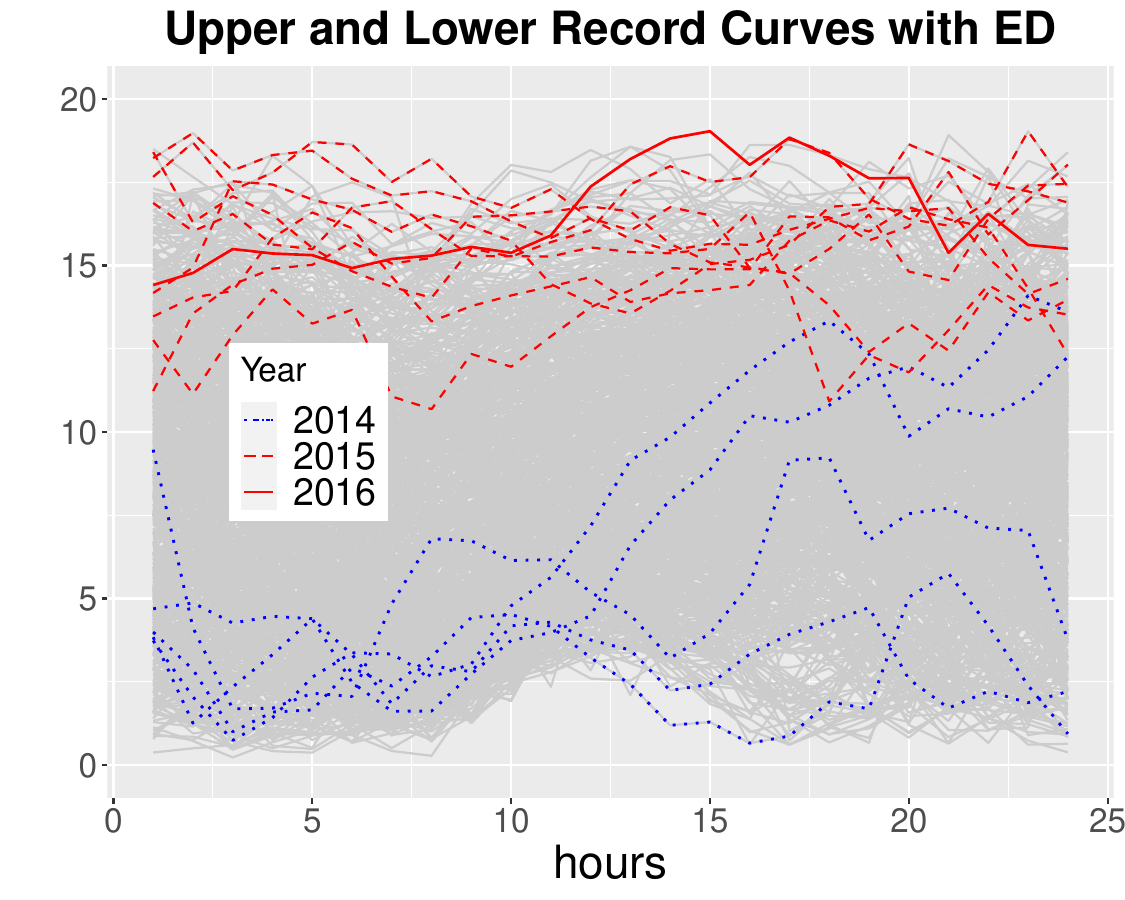}
  \includegraphics[scale=.3]{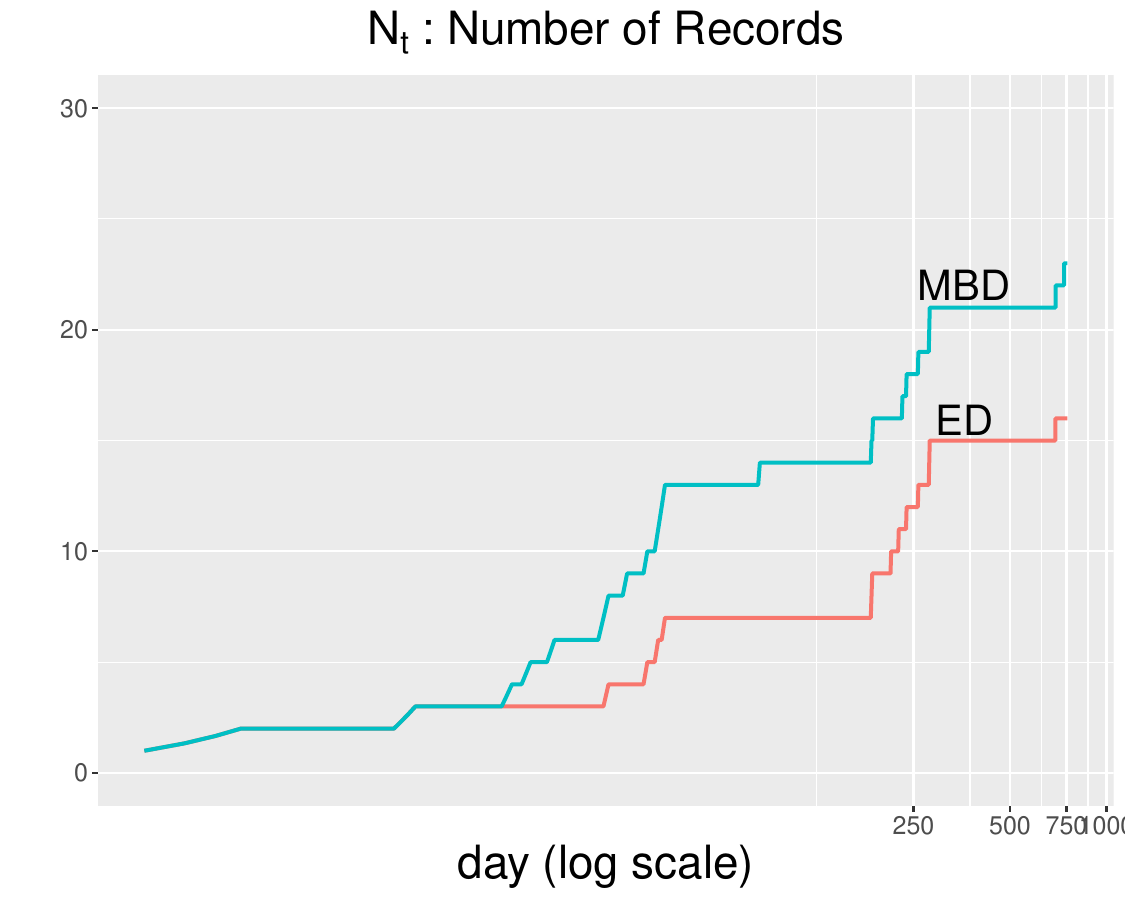}
\caption{ Functional records of daily wind speed at Yanbu, Saudi Arabia, from August $30$, $2014$ to September $22$, $2016$.  Left: Functional records obtained with MBD. Center: Functional records obtained with ED. Blue color indicates lower records and red color indicates upper records. Right: Trajectory of the number of functional records $N_{t}$, $t=2,\ldots, 755$. }
\label{WindData} 
\end{center}
\end{figure}

We exclude the two first curves that are functional records by definition. The dates in parentheses are those corresponding to ED.  We found that the functional records for $2014$ are: Sept. $5 (5),12,14,17,25,26 (26),30$, Oct. $5 (5),8 (8),9,10 (10)$, and Nov. $20$. The record curves for $2015$ are: Mar. $2 (2), (3),4, (30)$, Apr. $(11),17,24 (24)$, May $15 (15)$, and June $05 (05), 06 (06)$. The record curves for $2016$ are: July $(22),23$, and Sept. $5$. 
We plot the functional records on the left and in the  center of Figure \ref{WindData} using the depths MBD and ED, respectively; those curves that are not classified as records are indicated in gray. 
The record curves that correspond to September, $2014$, can be considered as part of the inherent variability of the functional process, so we do not include them in the plot. The lower functional records are indicated by the blue curves, and the upper functional records are indicated by the red curves. We indicate the corresponding year by using different line types, as follows: $2014$-dotted curves, $2015$-dashed curves, and $2016$-solid curves. We observe that all lower records are in $2014$ for both functional depths, whereas upper records are in $2015$ and $2016$. Thus, curves showing the lowest speeds are in Autumn, when the temperature starts to  slowly decrease. Most of the upper functional records were observed in Spring and Summer (except the last one with MBD, observed in September $2016$). Summer in Saudi Arabia brings sandstorms driven by Summer South winds. Therefore, it is reasonable to observe these extreme curves.

We can now infer the class of the underlying functional process.  On the right side of Figure \ref{WindData}, we present the trajectory of the corresponding $N_{t}$ process. We apply our RB-functional unit root test to the wind speed dataset. The test statistic value with MBD is $T_{n}=0.27$ and with ED is  $T_{n}=0.01$, both are smaller than the corresponding $5\%$ quantile $q_{0.05}=0.59$, even smaller than the $1\%$ quantile $q_{0.01}=0.34$. Thus, we have significant evidence against the stochastic trend, and conclude that the functional wind data do not have a unit root component. Therefore, the daily curves of the wind speed can be modeled with a stationary functional time series model. 

\subsection{Mortality rates in France}

This dataset consists of $n=191$ curves of annual mortality rates in France, from $1816$ to $2006$, for zero to $110$-years old individuals. However, we consider only up to $100$ years of age in order to avoid highly noisy measurements.  Each point of the curve $X_{i}(s)$ represents the total mortality rate, in year $i$, at age $s$. Our interest is to study the behavior of the rates, over the years,  taking into account all ages. By studying records, we analyze whether the new functional records over the years correspond to the natural randomness of the process, or if they indicate a decreasing trend. The data have been analyzed before by \cite{Hyndmanetal2007}  using a functional approach. They proposed to forecast the age-specific mortality rate by modeling the coefficients obtained by projecting the functional data to the corresponding robust functional principal components. They fitted an ARIMA model to the coefficients, but they did not report the estimated parameters. Evidence of a univariate unit root can be found if we fit the ARIMA model to the first coefficients, for the first eigenfunction. We therefore investigate if there is evidence of a functional unit root. In our analysis, we use the smoothed curves, as described  in  \cite{Hyndmanetal2007}.




We exclude the two first curves that are functional records by definition. The years in parentheses are those corresponding to ED.
We find that the years for the corresponding functional records are: $1818 (1818)$, $1819 (1819)$, $1821 (1821)$, $1832 (1832)$, $1845 (1845)$, $1862$, $1871 (1871)$, $(1872, 1884, 1887-1889, 1896)$, $1897 (1897)$, $(1910, 1912)$, $1913 (1913)$, $(1914, 1915,1918)$, $1920 (1920)$, $(1921,1922)$, $1923 (1923)$, $1924 (1924)$, $1927 (1927)$, $1930 (1930)$, $(1932)$,  $1933 (1933)$, $1934 (1934)$, $(1936)$, $1937 (1937)$, $1939$, $1946-1948 (1946-1948)$, $(1950-1955)$, $1958 (1958)$, $1959 (1959)$, $(1960)$, $1961 (1961)$, $1966 (1966)$, $(1974)$, $1975 (1975)$, $1977 (1977)$, $(1979)$, $1980 (1980)$, $1981$, $1984-1987 (1985-1987)$, $(1990, 1991)$, $1992-2006 (1992-2006)$. That is $49$ functional records in total with MBD, and $71$  functional records in total with ED.

\begin{figure}[!t]
\begin{center}
 \includegraphics[scale=.3]{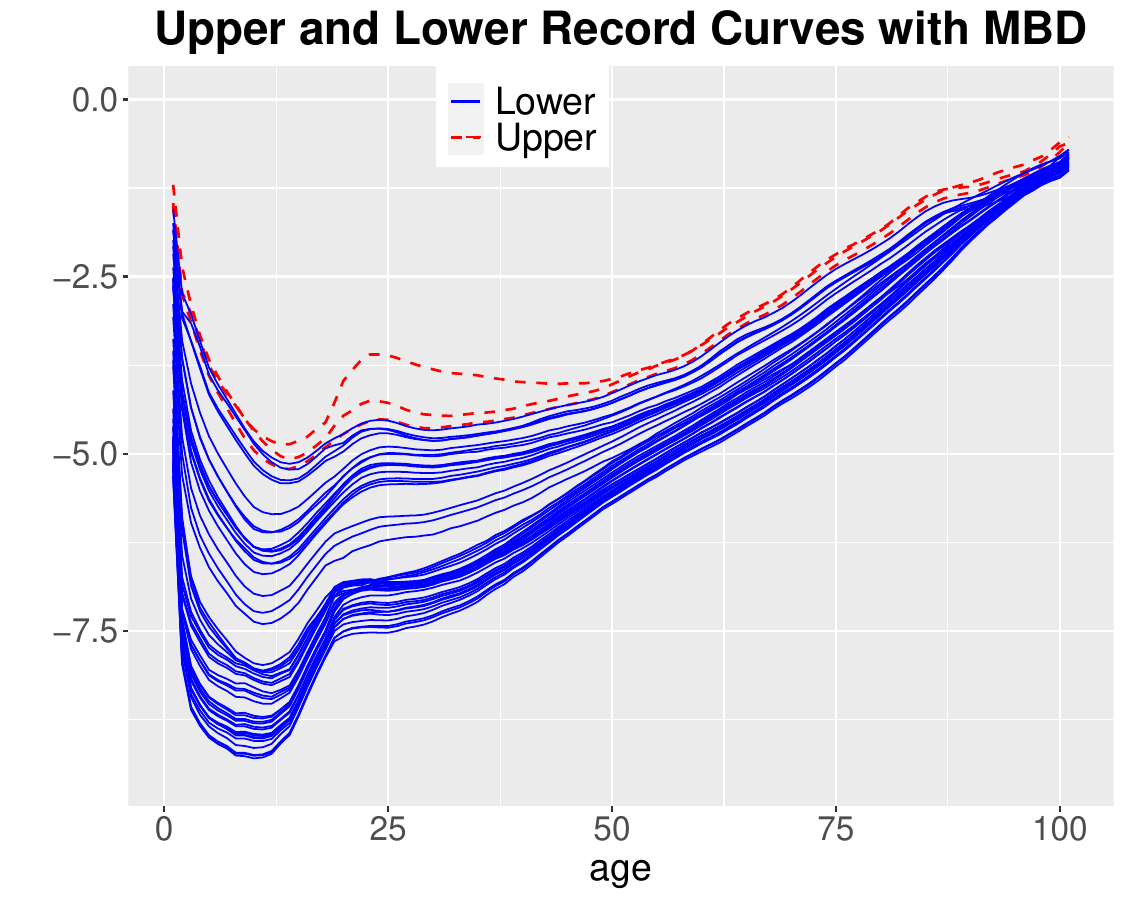}\hspace{-.3cm}
  \includegraphics[scale=.3]{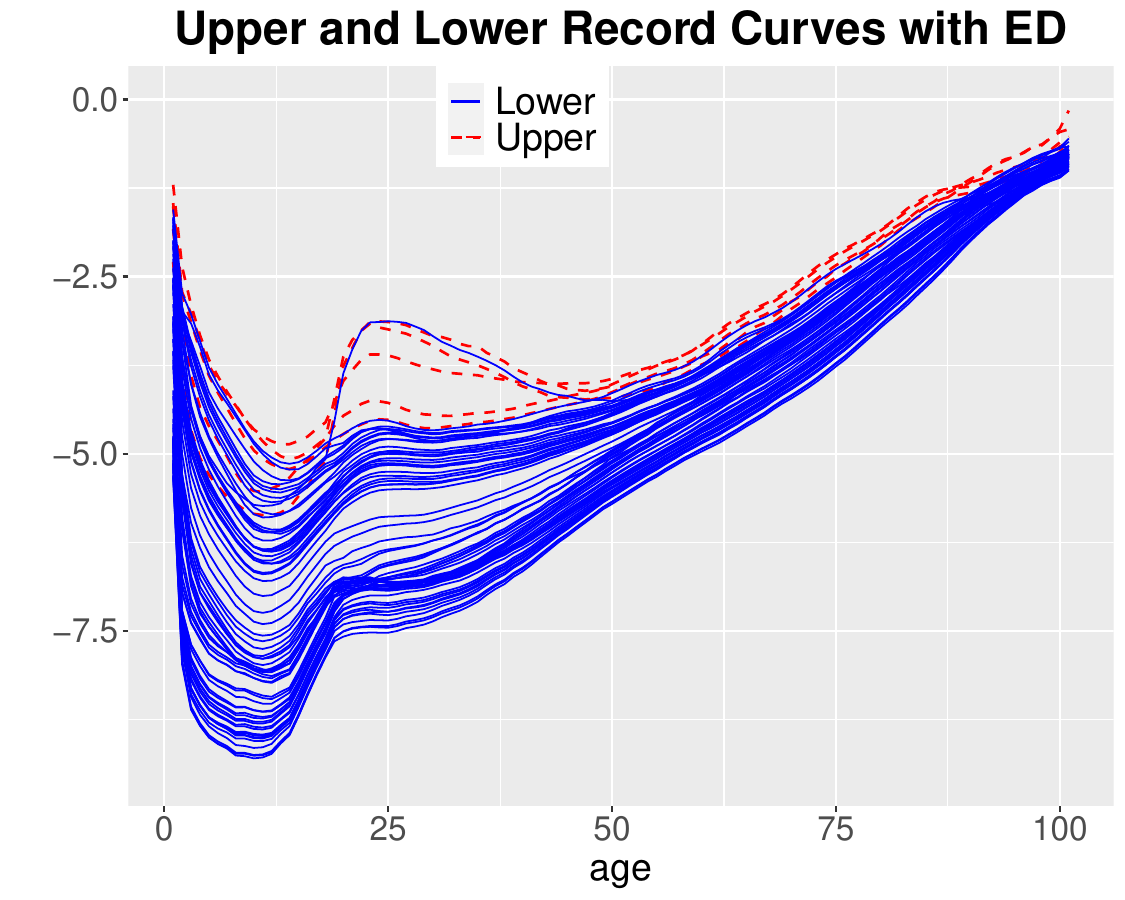}\hspace{-.1cm}
    \includegraphics[scale=.3]{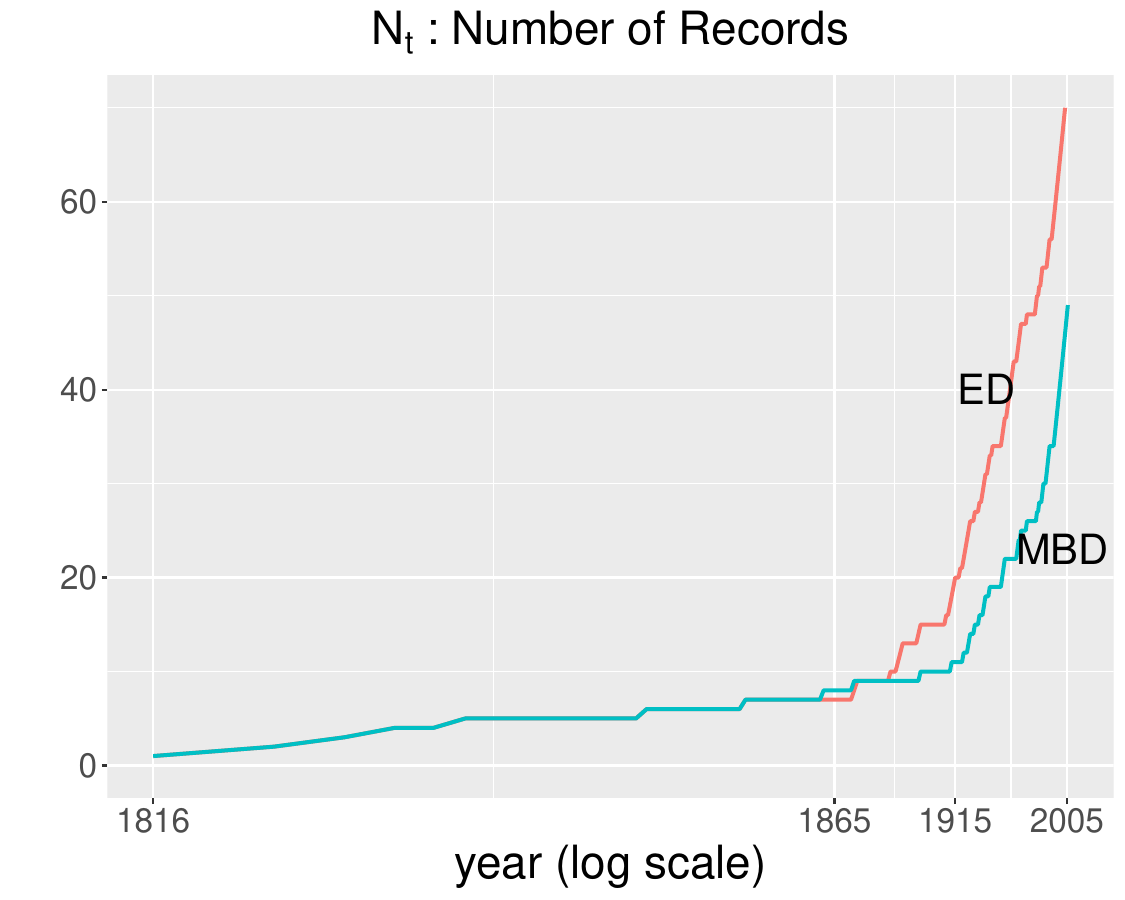}
\caption{Functional records of log mortality rates in France from $1816$ to $2006$, for zero to $100$ years of age. Left: Functional records obtained with MBD. Center: Functional records obtained with ED. Blue color indicates lower records and red color indicates upper records. Right: The trajectory of the number of functional records over time.}
\label{DataFrance} 
\end{center}
\end{figure}

Figure \ref{DataFrance} shows the functional records. We indicate the upper and lower records with different line types: upper functional records with a red dashed curve, and lower functional records with a blue solid curve. With MBD, we observe only three upper records that correspond to the years $1818, 1832$, and $1871$.  Whereas with ED, we observe two additional upper records; $1914$ and $1918$.  The rest of the records correspond to lower functional records. In particular, we observe that, after the last upper functional records in $1871$ ($1918$), a new functional record represents a lower mortality rate for almost all ages. This suggests the presence of a functional trend. 

Finally, we apply our RB-functional unit root test to the dataset. On the right side of Figure~\ref{DataFrance}, we show the trajectory of the corresponding $N_{t}$ process. The test statistic value with MBD is $T_{n}= 3.54$, and with ED is $T_{n}= 5.28$. The corresponding $5\%$ quantile from the asymptotic distribution under the null hypothesis is $q_{0.05}=0.59$. Therefore, we do not have any evidence against the  $I(1)$ functional process. Thus, to model this dataset, we must consider the existence of both a stochastic trend and a functional deterministic trend. This is consistent with the findings by \cite{Hyndmanetal2007} that take into consideration the ARIMA models for the basis coefficients. However, our approach is more general, as we do not consider any specific model.









\section{Discussion}  \label{Dis}

In this paper, we provided some statistical tools for functional time series. These tools are based on extending the record definition to functional data. We used a depth notion to rank curves and then be able to classify the extreme curves. The definition of a functional record considers jointly the upper and lower records. 
We showed that the counting process corresponding to the number of functional records grows at rate $\log t$, for stationary functional time series, and that it grows at rate $t^{1/2}$, for nonstationary functional time series. A simulation study showed that the asymptotic distribution of the number of records has a good approximation when the functional data are a functional random walk, even for small sample sizes. 

As a particular application of the extended functional record, we proposed a functional unit root test for a general definition of $I(1)$ functional processes. Using a Monte Carlo simulation study, we showed that the test performance is good for $I(0)$ and $I(1)$ functional processes. Our test is robust against structural changes for a moderate sample size. The unit root test based on functional records does not assume any model. In the data application, we found that the definition of functional records provides relevant and consistent information about extremes curves. In addition, it allows us to infer about the underlying process.


\section*{Data Availability Statement}
The wind speed data used in this paper will be available on request to the authors. The second data used in this paper are available from the \textit{demography} R package \url{https://cran.r-project.org/src/contrib/demography_1.22.tar.gz} \citep[][]{demography}. 

\section*{Appendix: Proofs}\label{sec:appendix}

\noindent \textbf{Proof of Proposition \ref{EstacionaryCase}:} 
First, we provide the proof for the i.i.d. case.  Let  $X_{1}, \ldots, X_{n}$  be a sequence of i.i.d functional random variables. Let $r_{n}(j)$ be the rank of the functional data $X_{j}$ among  $X_{1}, \ldots, X_{n}$, that is, $r_{n}(j)= 1+ \sum_{i=1, i\neq j}^{n} \mathds{1}\{ X_{j} \prec X_{i}\}$.  Since ``$\prec$'' is a strict ordering, we have that  $\{ r_{n} (1), \ldots, r_{n}(n)\}$ is a permutation of $1, \ldots, n$. Since $ r_{n} (1), \ldots, r_{n}(n)$ are clearly exchangeable, we conclude that $P\{ r_{n}(j)=k \}= 1/n$ for $k=1, \ldots, n$. Moreover, we have that $P\{r_{n} (1)=\sigma(1), \ldots, r_{n} (n)=\sigma(n) \}= 1/n!$, for any permutation $\sigma$ of values $1,\ldots, n$.  Now let  $R^{u}_{j}$ be a Bernoulli random variable with  $P(R^{u}_{j}=1)= P(r_{j}(j)= j)= 1/j$ and variance $(1- 1/j)/j$. Thus, to prove that $P(\lim_{t\to \infty} \frac{N_{t}^{u}}{\log t} =1)=1$, we use the Kolmogorov convergence criterion \cite[Chap~6]{Gut2013}, and the fact  that $N^{u}_{t}=\sum_{j=1}^{t} R^{u}_{j}$. 

We observe that $\sum_{j=1}^{\infty} \mathrm{Var} \left( \frac{R^{u}_{j}- 1/j}{\log j}\right) \leq \sum_{j=1}^{\infty} \frac{1}{j (\log i)^{2}}<\infty$. Then, we have that $\sum_{j=1}^{\infty}  \frac{R^{u}_{j}- 1/j}{\log j} $ converges with probability one.  Therefore $\frac{1}{\log n} \sum_{j=1}^{n} (R^{u}_{j}- 1/j) \overset{a.s.}{\to}  0$. Finally, we note that $\frac{1}{\log t}  \sum_{j=1}^{t} R^{u}_{j}- 1/j =\frac{1}{\log t}\left( N_{t}^{u} - \sum_{j=1}^{t}  1/j\right)$, and therefore
$$\frac{N_{t}^{u} }{\log t} -1 = \frac{N_{t}^{u}-  \sum_{j=1}^{t}  1/j  }{\log t} + \frac{ \sum_{j=1}^{t}  1/j -\log t }{\log t}\to 0,$$
where in the second term we used that $ \sum_{j=1}^{t}  1/j -\log t \to \gamma$ as $t\to \infty$, with $\gamma$ the Euler’s constant. This  proves the proposition for the i.i.d. case. 

To prove the stationary case, one can use similar ideas as in \cite{Leadbetter-and-Rootzen88}. That is, one assumes that the functional time series $\{X_{i}\}$ can be approximated with  a sequence $\{X_{i}^{(m)}\}$ that is $m$-dependent and strictly stationary. This concept is known as  $L^{p}-m$-approximable \citep[][]{HormanKokoszka}.  Specifically, $\{X_{i}\}$ is such that 
$\sum_{i=1}^{\infty} (\mathbb{E} \|X_{i}- X_{i}^{(m)} \| )^{1/p}<\infty$. This assumption is not restrictive, and most of the commonly used time series models (scalar, vector, or functional) satisfy this assumption. Hence, the autocovariances $C_{h}$ of $X_{i}$  decrease to zero exponentially as $h\to \infty$, and the assumption $\log (h) \|C_{h} \|_{\mathcal{S}}\to 0$ is satisfied. Under this assumption,  $\{X_{i}\}$ can be replaced by the $m$-dependent sequences $\{X_{i}^{(m)}\}$. This reduces the proof to proving the result for the $m$-dependent sequences. 

Without loss of generality,  assume $n=k m $, with $m$ fixed.  The next step is to divide the set $\{1, \ldots, n\}$ into $m$ subsets of length $k$. Now, we have that $\{X_{1}, \ldots, X_{n}\}= \cup_{j=1}^{m} \{ X_{j+im},\, i=0, 2, \ldots, k\} $, and for each $k$,  $\{ X_{j+im},\, i=0, 2, \ldots, k\}$ is a sequence  of independent random variables. Let $X_{n}$ be an upper record at time $n$, i.e., 
$\mathrm{fD}(X_{n}; \mathbf{X}_{1:n})\leq \mathrm{fD}(X_{L^{u}(N^{u}_{n-1})}; \mathbf{X}_{1:n})$,  then, it is not difficult to see that  $\mathrm{fD}(X_{n}; \mathbf{X}_{A}) < \mathrm{fD}(X_{j};  \mathbf{X}_{A})$, for all  $j \in A$, $n\neq j$, and $A$ any subset of $\{1, \ldots, n\}$ containing $n$ (the equality would only happen if $L^{u}(N^{u}_{n-1}) \in A$). That is, if  $X_{n}$ is an upper record in $\{X_{1}, \ldots, X_{n}\}$, then $X_{n}$ is also an upper record in any subset of $\{X_{1}, \ldots, X_{n}\}$ containing $X_{n}$. Now, take $A_{j}$ as $\{j+im, i=0,1,\ldots, k \}$, $j=1, \ldots, m$, then we have that the number of upper records $N_{n}^{u}$ in $\{X_{1}, \ldots, X_{n}\}$ is smaller than the total number of records  in $ \mathbf{X}_{A_{1}}, \ldots,  \mathbf{X}_{A_{m}}$. Consequently $\frac{ \tilde{N}_{n}^{u}}{\log n}\leq \frac{N_{n}^{u}}{ \log n} \leq \frac{m \tilde{N}_{n}^{u} }{\log n}$ for $n\gg 0$, where $\tilde{N}_{n}^{u} $ is the number of upper functional records corresponding to the i.i.d. case. Therefore, $\lim_{n\to \infty} N_{n}^{u}/\log n = O(1)$, since $\lim_{n\to \infty} m \tilde{N}_{n}^{u}/\log n  = O(1)$. 
\hfill $\Box$
\vspace{.3cm}

Before proving Prepositions \ref{LimDistiUR} and \ref{LimDist3}, we will first notice that the order statistic induced by a functional depth has a  Markov property. Let $X_{0},X_{1}, \ldots, X_{n}$ be the functional time series starting at time $0$, and $N_{n}$ the number of records until time $n$.
For $j\geq 2$, let $\tilde{R}_{j}$ be the indicator of $X_{j}$ being a functional record but using only three curves. Specifically, let $\{\boldsymbol{\mathcal{X}}_{i},\, i\geq 1\}$ be a bivariate functional process defined as follows: $\boldsymbol{\mathcal{X}}_{1}= (X_{0}, X_{1})^{T}$, and for $i\geq 2$, $\boldsymbol{\mathcal{X}}_{i}= ( \mathcal{X}_{i,1},\mathcal{X}_{i,2})^{T}$ where $\{ \mathcal{X}_{i,1}, \mathcal{X}_{i,2}\}= \{\mbox{the two most extreme curves among } \{\mathcal{X}_{i-1,1}, \mathcal{X}_{i-1,2}, X_{i}\} \}$. The bivariate process $\boldsymbol{\mathcal{X}}_{i}$ is a constant process until $X_{i}$ is not the deepest curve in the three elements set. Then, $\tilde{R}_{j}:= \mathds{1}\{X_{j}\in \{\mathcal{X}_{j,1}, \mathcal{X}_{j,2} \} \}$.  

Now, we will prove $\sum_{j=1}^{n}R_{j} = \sum_{j=1}^{n}\tilde{R}_{j}$. Clearly $\tilde{R}_{2}=R_{2}$. Now, for $j\geq 3$. Assume that $X_{j}$ is a functional record at time $j$, and without loss of generality, assume it is an upper functional record, i.e., $\mathrm{fD}(X_{j}; \mathbf{X}_{1:j})\leq \mathrm{fD}(X_{L^{u}(N^{u}_{j-1})}; \mathbf{X}_{1:j})$. If there are no ties, i.e., $\mathrm{fD}(X_{j}; \mathbf{X}_{1:j})< \mathrm{fD}(X_{L^{u}(N^{u}_{j-1})}; \mathbf{X}_{1:j})$, then $\mathrm{fD}(X_{j}; \boldsymbol{\mathcal{X}}_{j-1}, X_{j})<\max\{ \mathrm{fD}(X_{L^{u}(N^{u}_{j-1})}; \boldsymbol{\mathcal{X}}_{j-1}, X_{j} ),  \mathrm{fD}(X_{L^{l}(N^{u}_{j-1})}; \boldsymbol{\mathcal{X}}_{j-1}, X_{j}) \}$. That is, $R_{j}=1$ implies $\tilde{R}_{j}=1$. If there are ties, i.e., $\mathrm{fD}(X_{j}; \mathbf{X}_{1:j})=\mathrm{fD}(X_{L^{u}(N^{u}_{j-1})}; \mathbf{X}_{1:j})$, then we have that $\mathrm{fD}(X_{j}; \boldsymbol{\mathcal{X}}_{j-1}, X_{j})\leq \max\{ \mathrm{fD}(X_{L^{u}(N^{u}_{j-1})}; \boldsymbol{\mathcal{X}}_{j-1}, X_{j}),  \mathrm{fD}(X_{L^{l}(N^{u}_{j-1})}; \boldsymbol{\mathcal{X}}_{j-1}, X_{j}) \}$, in other case there is another curve $X_{i_{0}}$, with $1\leq i_{0}< j-1$ and $i_{0}\neq L^{u}(N^{u}_{j-1}), L^{l}(N^{u}_{j-1})$, such that $ \mathrm{fD}(X_{i_{0}}; \boldsymbol{\mathcal{X}}_{j-1}, X_{j}) \leq \mathrm{fD}(X_{j}; \boldsymbol{\mathcal{X}}_{j-1}, X_{j})$, that is, $\mathrm{fD}(X_{j};\mathbf{X}_{1:j} ) \leq \mathrm{fD}(X_{i_{0}}; \mathbf{X}_{1:j})\leq \mathrm{fD}(X_{L^{u}(N^{u}_{j-1})}; \mathbf{X}_{1:j})$, and as a consequence $X_{i_{0}}\sim X_{L^{u}(N^{u}_{j-1})} \sim X_{j}$. By Assumption \ref{A2} this event as probability zero. Thus, $R_{j}=1$ implies $\tilde{R}_{j}=1$ when ties occur. Therefore $\sum_{j=1}^{n}R_{j} \leq \sum_{j=1}^{n}\tilde{R}_{j}$. Now, assume that $X_{j}$ is not a functional record at time $j$. Then we have $\mathrm{fD}(X_{j}; \mathbf{X}_{1:j})>  \mathrm{fD}(X_{L^{u}(N^{u}_{j})},\mathbf{X}_{1:j})$ and  $\mathrm{fD}(X_{j}; \mathbf{X}_{1:j})>  \mathrm{fD}(X_{L^{l}(N^{l}_{j})}; \mathbf{X}_{1:j})$, otherwise it would be a functional record. Since $L^{u}(N^{u}_{j})= L^{u}(N^{u}_{j-1})$ and $L^{l}(N^{l}_{j})= L^{l}(N^{l}_{j-1})$, we have that $\mathrm{fD}(X_{j};  \boldsymbol{\mathcal{X}}_{j-1}, X_{j} )> \max\{ \mathrm{fD}(X_{L^{u}(N^{u}_{j-1})};  \boldsymbol{\mathcal{X}}_{j-1}, X_{j}),  \mathrm{fD}(X_{L^{l}(N^{l}_{j-1})};  \boldsymbol{\mathcal{X}}_{j-1}, X_{j}) \}$, i.e., $X_{j}$ is deeper than $X_{L^{u}(N^{u}_{j-1})}$ and $X_{L^{l}(N^{l}_{j-1})}$ in $\{X_{L^{u}(N^{u}_{j-1})}, X_{L^{l}(N^{l}_{j-1})}, X_{j}\}$. As a consequence $R_{j}=0$ implies $\tilde{R}_{j}=0$, thus  $\sum_{j=1}^{n}R_{j} \geq \sum_{j=1}^{n}\tilde{R}_{j}$. Therefore $\sum_{j=1}^{n}R_{j} = \sum_{j=1}^{n}\tilde{R}_{j}$ with probability one. 
\vspace{.3cm}

\noindent \textbf{Proof of Proposition \ref{LimDistiUR}:} 

To prove Proposition \ref{LimDistiUR}, we use $\tilde{R}_{j}$ instead of $R_{j}$.  
 We consider the bivariate functional time series defined by $\mathbf{X}_{1}=( X_{1}, X_{0})^{T}, \mathbf{X}_{2}=( X_{2}, X_{1})^{T}, \ldots, \mathbf{X}_{n}=( X_{n}, X_{n-1})^{T}$. We observe that the components of $\mathbf{X}_{1}$ are record curves by definition. We denote by $\boldsymbol{\tau}=\{\tau_{1}, \ldots, \tau_{N_{n}^{u}}\}^{T}$ where $\tau_{1}=1$, for $j=2,\ldots, N_{n}^{u}-1$, $\tau_{j}$ is the time interval between the upper record $j$ and $j+1$, and $\tau_{N_{n}^{u}}=n- \sum_{j=1}^{N_{n}^{u}-1} \tau_{j}$.

Let $\mathbf{Y}^{u}_{1}=\boldsymbol{\mathcal{X}}_{1}$, \ldots, $\mathbf{Y}^{u}_{\tau_{1}+\tau_{2}-1}=\boldsymbol{\mathcal{X}}_{1}, \mathbf{Y}^{u}_{\tau_{1}+\tau_{2}}=\boldsymbol{\mathcal{X}}_{\tau_{1}+\tau_{2}}, \ldots, \mathbf{Y}^{u}_{\tau_{1}+\tau_{2}+\tau_{3}-1}=\boldsymbol{\mathcal{X}}_{\tau_{1}+\tau_{2}}$, $\mathbf{Y}^{u}_{\tau_{1}+\tau_{2}+\tau_{3}}= \boldsymbol{\mathcal{X}}_{\tau_{1}+\tau_{2}+\tau_{3}}, \ldots$. The bivariate process $\{\mathbf{Y}^{u}_{j}\}$ contains functional record curves, and it ``jumps'' when a new upper functional record is observed. 
We consider the joint distribution $P(\boldsymbol{\tau}, N_{n}^{u} )$ of record times $\boldsymbol{\tau}$ and number  of records $N_{n}^{u}$. 

Since $X_{i}$ is a functional random walk, it has the Markov property, and by using the translation invariance with respect to the initial curve, the probability of $X_{i}$ being an upper record only depends on $\mathbf{Y}^{u}_{L(N_{i}^{u}-1)}$ instead of all the past. For a number of curves $n$,
\begin{equation}\label{JointProb}
P(\boldsymbol{\tau}, N_{n}^{u} )= p(\tau_{1} | \mathbf{Y}^{u}_{\tau_{1}} )p(\tau_{2}| \mathbf{Y}^{u}_{\tau_{1}+\tau_{2}})\cdots p(\tau_{N^{u}_{n}-1}|  \mathbf{Y}^{u}_{L(N^{u}_{n}-1)}\ ) q( \tau_{N^{u}_{n}}| \mathbf{Y}^{u}_{L(N^{u}_{n}-1)}),
\end{equation}
where $p(t | \mathbf{Y} )=P(  \mathbf{X}_{2}\prec \mathbf{Y},  \mathbf{X}_{3} \prec \mathbf{Y}, \ldots, \mathbf{X}_{t -1} \prec \mathbf{Y}, \mathbf{X}_{t} \succ \mathbf{Y} \,|\, \mathbf{Y} ) $ and  $q(t | \mathbf{Y})=P(  \mathbf{X}_{2}\prec \mathbf{Y},  \mathbf{X}_{3} \prec \mathbf{Y}, \ldots, \mathbf{X}_{t}  \prec \mathbf{Y} \,| \,\mathbf{Y} ) $. Because of the translation invariance of $\{X_{i}\}$ with respect to the initial curve, $p(t | \mathbf{Y} )$ and $q( t | \mathbf{Y} )$ do not depend on $\mathbf{Y}$. From the way we write the bivariate process, we obtain that $p(t)$ and $q(t)$ correspond to a  general renewal process. Following \cite{Feller71}, Chap $7$,
 the generating function of $q(t)$ is 
$$\tilde{q}(z)= \sum_{t=0}^{\infty} q(t) z^{t}= \exp \left\{ \sum_{t\geq 1} \frac{z^{t}}{t} P(X_{t} \leq 0) \right\}. $$
On the other hand, if we consider the generating function of \eqref{JointProb} and take the summation over all possible values of components of $ \boldsymbol{\tau}$ and all possibles sample sizes $n$, we find that, by using $p(t)= q(t-1)- q(t)$:
\begin{equation} \label{GMJointP}
\sum_{n=0}^{\infty} P (N_{n}^{u}=k) z^{n}= \{1 - (1-z) \tilde{q}(z)\}^{k-1} \tilde{q}(z),\quad k>1.
\end{equation}

Since the distribution of the functional white noise $\varepsilon_{0}$ is assumed to be symmetric, then $P(X_{i} \leq 0)=1/2$. Therefore, using that $\sum_{j\geq 1} z^{j} /j =- \log (1-z) $ we obtain $\tilde{q}(z)= \frac{1}{\sqrt{1-z}}$, and then $\sum_{n=0}^{\infty} P (N_{n}^{u}=k) z^{n}=(1-\sqrt{1-z})^{k-1} / \sqrt{1-z}. $ Now, by expanding the right side of this equation in powers of $z$, 
we obtain that $P (N_{n}^{u}=k) $ has the form $\binom{2n -k+1}{n}2^{-2n +k-1}.$ Finally, taking the limit as $n \to \infty$, we obtain that 
$ N_{n}^{u} / \sqrt{n}   \overset{d}{\longrightarrow}   G_{1}$, where $G_{1}$ has density $g_{1}(x)= \frac{1}{\sqrt{\pi}} \exp(-x^{2}/4) $ for $x\geq 0$.

\hfill $\Box$

\noindent \textbf{Proof of Proposition \ref{LimDist3}:} Let $\{X_{i}\}$ be an $I(1)$ functional process, then $X_{i}$ can be written as 
$X_{i}= Z_{0}+ \Psi\left(\sum_{j=1}^{i} \varepsilon_{j} \right)+ \nu_{i}$. 
We are interested in the distribution of the random variable $N^{u}_{n} /\sqrt{n}$, when $n\to \infty$. Notice that the functional records are computed on the Hilbert space $\mathcal{H}$, and this space $\mathcal{H}$ can be written as direct sum of the subspaces $(\mathrm{ker} \, \Lambda)^{\bot}$ and $\mathrm{ker} \, \Lambda$: $\mathcal{H}= (\mathrm{ker} \, \Lambda)^{\bot} \oplus \mathrm{ker} \, \Lambda$. To prove the result, we use this direct sum. The intuition behind the proof is that  $X_{i}$ is driven by the variance of the process  $\Psi\left(\sum_{j=1}^{i} \varepsilon_{j} \right)$. Thus, the number of records corresponding to $\{X_{i}\}$ has the same asymptotic distribution as the one corresponding to the functional process $\Psi\left(\sum_{j=1}^{i} \varepsilon_{j} \right)$ on the closure subspace defined by $\mathrm{ran}\, \Psi $.

First, let $v\notin \mathrm{ker}\, \Psi^{*} = \mathrm{ker} \, \Lambda$, then $\langle X_{i}, v \rangle =\langle Z_{0}, v \rangle + \langle \Psi\left(\sum_{j=1}^{i} \varepsilon_{j} \right) , v \rangle + \langle \nu_{i}, v \rangle= Z_{0}^{v} + \sum_{j=1}^{i}  \langle \varepsilon_{j}, \Psi^{*} v \rangle  + \nu_{i}^{v} $, where $ Z_{0}^{v} $ is a scalar, and $\nu_{i}^{v}$ is a scalar stationary time series. Also, $\langle \varepsilon_{j}, \Psi^{*} v \rangle$ is an independent random variable with non-degenerate distribution, since $\Psi^{*} v\neq 0$.  Therefore,  $\{\langle X_{i}, v \rangle\}$ is a random walk with stationary errors $\nu_{i}^{v}$, for all $v\notin \mathrm{ker}\, \Psi^{*}$. We first prove that  $N_{t}^{u}/ \sqrt{t}$ on $(\mathrm{ker}\, \Lambda)^{\bot}$ has the same asymptotic distribution as the corresponding one for functional random walk. For that, we use the notation $N_{n}^{u} (X)$ to emphasize that the upper records correspond to the process $X$.  Let $P_{1}$ be the oblique projection on $(\mathrm{ker}\, \Lambda)^{\bot}$ along $\mathrm{ker}\, \Lambda$, and let $N_{n}^{u} (\mathrm{rw})$ denote the number of upper records of a pure random walk (meaning $\nu_{i} \equiv 0$).   Then,  the number of upper records on  $(\mathrm{ker}\, \Lambda)^{\bot}$ is such that $\frac{N_{n}^{u} (P_{1}(\mathrm{rw}))}{\sqrt{n}} \leq \frac{ N_{n}^{u}(P_{1}(X))}{\sqrt{n}} \leq \frac{N_{n}^{u}( P_{1} (\mathrm{rw}))}{\sqrt{n}} + \frac{N_{n}^{u} (P_{1} (\nu))}{\sqrt{n}} $. Consequently,  $\frac{N_{n}^{u} (P_{1}(\mathrm{rw}))}{\sqrt{n}} = \frac{ N_{n}^{u}(P_{1}(X))}{\sqrt{n}}$  as $ n \to \infty$, since $ \frac{N_{n}^{u} (P_{1} (\nu))}{\sqrt{n}} \approx \frac{\log n}{\sqrt{n}}\to 0$. Therefore, the random variable $N_{n}^{u} / \sqrt{n}$ on $(\mathrm{ker}\, \Lambda)^{\bot}$ has the same asymptotic distribution as upper records of functional random walk.
 
Now, consider the space $\mathrm{ker} \, \Lambda$. Let $v \in \mathrm{ker} \, \Lambda$, then $\langle X_{i}, v \rangle =\langle Z_{0}, v \rangle + \sum_{j=1}^{i}  \langle \varepsilon_{j}, \Psi^{*} (v) \rangle + \langle \nu_{i}, v \rangle= Z_{0}^{v}  + \nu_{i}^{v} $, since $ \Psi^{*} (v)=0 $.  Therefore, $\{\langle X_{i}, v \rangle\}$ is stationary for all $v \in \mathrm{ker} \, \Lambda$, since $\nu_{i}^{v}$ is stationary.  Let $P_{2}$ be the oblique projection on $\mathrm{ker}\, \Lambda$ along $(\mathrm{ker}\, \Lambda)^{\bot}$. From Proposition  \ref{EstacionaryCase}, we conclude that $N_{n}^{u} (P_{2} (X_{n})) = O(\log n)$.

Finally,  we observe that the random variable $N_{n}^{u}/ \sqrt{n}$ can be bounded as
$$\frac{N_{n}^{u}(P_{1}(X))}{ \sqrt{n}} \leq \frac{N_{n}^{u}( X)}{ \sqrt{n}} \leq \frac{N_{n}^{u}( P_{1}(X))}{ \sqrt{n}} + \frac{N_{n}^{u}(P_{2} (X))}{ \sqrt{n}}.$$
Therefore, we conclude that $\frac{N_{n}^{u}(P_{1}(X))}{ \sqrt{n}} = \frac{N_{n}^{u}( X)}{ \sqrt{n}} $ as $n\to \infty$. 
 \hfill $\Box$

\noindent \textbf{Proof of Corollary \ref{AD}:} 

Similarly to the random variable $N_{n}^{u}$, we have that  $N_{n}^{l}/\sqrt{n} \overset{d}{\longrightarrow}  G_{1}$, where $N_{n}^{l}$ is the corresponding counting process for the lower functional records.
For renewal processes, we know that the asymptotic joint distribution of $(N_{n}^{u}/\sqrt{n}, N_{n}^{l}/\sqrt{n} )$ is equal to the joint distribution of $(|W(1)|, l(0,1) )$, where $W(s)$ is the Brownian motion and $l(0,1)$ is the local time of the Brownian motion at zero,  evaluated at one. On the other hand $(|W(u)|, l(0,u) )$ and $( \max_{0\leq s \leq u} W(s) - W(u), \max_{0\leq s \leq u} W(s) )$ have the same joint density $f_{0}(x,y)$. Using that the joint density function of $( \max_{0\leq s \leq u} W(s),  W(u)  )$ is $f(m,w)= \sqrt{\frac{2}{\pi u^{3/2}} }(2m-w)\exp\{- (2m-w)^{2} / 2u  \} $, we obtain that $f_{0}(x,y)=\sqrt{\frac{2}{\pi u^{3/2}} }(x+y)\exp\{- (x+y)^{2} / 2u  \} $. Using a bivariate transformation of the random variables, we obtain that the asymptotic distribution of $(N_{n}^{u} + N_{n}^{l})/\sqrt{n}$ has density 
$g_{2}(x)=\sqrt{\frac{2}{\pi}}\,x^{2}\exp (- x^{2}/2 ),\,\, x \geq 0. $
\hfill $\Box$

\baselineskip=16pt

\bibliographystyle{chicago}

\end{document}